\documentclass[printer]{aa}
\usepackage{txfonts}
\usepackage{graphicx}
\usepackage{url}
\usepackage{hyperref}
\usepackage{psfrag}
\DeclareGraphicsExtensions{.pdf,.png,.jpg,.ps,.eps}
\usepackage{natbib}
\bibpunct{(}{)}{;}{a}{}{,} 

\newcommand{\msun}{\mbox{M$_{\odot}$}}
\newcommand{\rsun}{\mbox{R$_{\odot}$}}
\newcommand{\lsun}{\mbox{L$_{\odot}$}}

\begin{document}
\title{Optical/near-infrared observations of the Fried Egg Nebula}
\subtitle{Multiple shell ejections on a 100 yr timescale from a massive yellow hypergiant}

\author{E.~Koumpia\inst{\ref{inst1},\ref{inst14}}, R.~D.~Oudmaijer\inst{\ref{inst1}}, V.~Graham\inst{\ref{inst1}}, G.~Banyard\inst{\ref{inst7}}, J.~H.~Black\inst{\ref{inst4}}, C.~Wichittanakom\inst{\ref{inst1},\ref{inst10}}, K.~M. Ababakr \inst{\ref{inst11}}, W.-J.~de Wit\inst{\ref{inst2}}, F. Millour\inst{\ref{inst3}}, E. Lagadec\inst{\ref{inst3}}, S.~Muller\inst{\ref{inst4}}, N.~L.~J.~Cox\inst{\ref{inst5}}, A. Zijlstra\inst{\ref{inst6},\ref{inst13}}, H.~van Winckel\inst{\ref{inst7}}, M.~Hillen \inst{\ref{inst7}}, R.~Szczerba\inst{\ref{inst8}}, J.~S. Vink\inst{\ref{inst12}}, S.~H.~J.~Wallstr{\"o}m\inst{\ref{inst7}}} 

\institute{School of Physics \& Astronomy, University of Leeds, Woodhouse Lane, LS2 9JT Leeds, UK\label{inst1}
\and
ESO Vitacura, Alonso de C{\'o}rdova 3107 Vitacura, Casilla 19001 Santiago de Chile, Chile\label{inst2}
\and 
Observatoire de la Cote d’Azur Nice, 96 Boulevard de l'Observatoire, 06300 Nice, France\label{inst3}
\and
Department of Space, Earth, and Environment, Chalmers University of Technology, Onsala Space Observatory, 43992 Onsala, Sweden\label{inst4}
\and ACRI-ST, 260 Route du Pin Montard, BP234, 06904 Sophia-Antipolis, France\label{inst5}
\and
Jodrell Bank Centre for Astrophysics, Alan Turing Building, The University of Manchester, Oxford Road, Manchester M139PL, UK\label{inst6}
\and
Instituut voor Sterrenkunde (IvS), KU Leuven, Celestijnenlaan 200D, 3001, Leuven, Belgium\label{inst7}
\and
Nicolaus Copernicus Astronomical Centre, PAS, ul. Rabia{\'n}ska 8, 87-100 Toru{\'n}, Poland\label{inst8}
\and
Department of Physics, Faculty of Science and Technology, Thammasat University, Rangsit Campus, Pathum Thani 12120, Thailand\label{inst10}
\and 
Erbil Polytechnic University, Kirkuk Road, Erbil, Iraq\label{inst11}
\and 
Armagh Observatory and Planetarium, College Hill, Armagh BT61 9DG, Ireland\label{inst12}
\and
Laboratory for Space Research, University of Hong Kong\label{inst13}
\and \email{ev.koumpia@gmail.com}\label{inst14}
}
\date{Received date/Accepted date}

\abstract
{The fate of a massive star during the latest stages of its evolution is highly dependent on its mass-loss rate/geometry and therefore knowing the geometry of the circumstellar material close to the star and its surroundings is crucial.}{We aim to give an insight regarding the nature (i.e. geometry, rates) of mass-loss episodes, and in particular the connection between the observed asymmetries due to the mass lost in a fast wind, or during a previous, prodigious mass-losing phase. In this context, yellow hypergiants offer a good opportunity to study mass-loss events.}{We analyse a large set of optical/near-infrared data, in spectroscopic and photometric (X-shooter/VLT), spectropolarimetric (ISIS/WHT), and interferometric GRAVITY-AMBER/VLTI) modes, toward the yellow hypergiant IRAS 17163-3907. We use the optical observations to determine its spectral type and we present the first model independent reconstructed images of IRAS 17163-3907 at these wavelengths tracing milli-arcsecond scales. Lastly, we apply a 2D radiative transfer model to fit the dereddened photometry and the radial profiles of published diffraction limited VISIR images at 8.59 $\mu$m, 11.85 $\mu$m and 12.81 $\mu$m simultaneously, adopting a revised distance determination using the new Gaia measurements (DR2).}{The interferometric observables around the 2 $\mu$m window towards IRAS 17163-3907 show that the Br$\gamma$ emission appears to be more extended and asymmetric than the \ion{Na}{i} and the continuum emission. In addition, Br$\gamma$ shows variability in a time interval of four months, not seen towards \ion{Na}{i}. Lastly, in addition to the two known shells surrounding IRAS 17163-3907 we report on the existence of a third hot inner shell with a maximum dynamical age of only 30 yr.}{The interpretation of the presence of \ion{Na}{i} emission at closer distances to the star compared to Br$\gamma$ has been a challenge in various studies. To address this, we examine several scenarios, and we argue that the presence of a pseudophotosphere which was traditionally considered to be the prominent explanation is not needed, but it is rather an optical depth effect. The three observed distinct mass-loss episodes are characterised by different mass-loss rates and can inform the theories on mass-loss mechanisms, which is a topic still under debate both in theory and observations. We discuss these in the context of photospheric pulsations and wind bi-stability mechanisms.
}


\titlerunning{Optical/near-infrared observations of the Fried Egg Nebula} 
\authorrunning{Koumpia et al.} 
 \maketitle


 \section{Introduction}
\label{intro}

Many important open issues related to the final stages of stellar evolution can only be addressed once we have knowledge of the geometry of the circumstellar material close to the star. For massive stars \citep[$8-30~\msun$;][]{Wang1998} which eventually explode as core-collapse supernovae (SN), the problem in understanding the mass-loss rate and mass-loss geometry is particularly acute \citep{Heger1998}. Mass-loss events impact the angular momentum evolution and final mass, and thus fate of the individual massive star, and they create the circumstellar environment that the SN ejecta may interact with. Following core-collapse explosion, the progenitor's mass-loss history continues to be relevant because of its repercussion on the light-curve morphology and probably also on the observed aspherical SN remnants \citep[e.g.][]{Patat2011,Moriya2014}. The processes responsible for shaping these spatial asymmetries are largely unknown.

This paper aims to give an insight regarding the physical structure, geometry, mass-loss episodes, and properties of the host star, which are amongst the main unknowns in theoretical models on stellar evolution. A suitable class of stars to study mass-loss events are the post-red supergiants (post-RSG), but only few such objects are known \citep{Oudmaijer2009}. The yellow (A to K-type) hypergiants (YHGs), such as IRAS 17163-3907, IRC+14020 and HD179821, that show evidence of circumstellar dust and high mass-loss rates, are great post-RSG candidates, and therefore laboratories to study the mass-loss events that take place during the post-RSG volution. 

In this study we focus on IRAS 17163-3907 (hereafter IRAS~17163), also known as Hen 3-1379, and WRAY 15-1676, which is one of the brightest infrared sources in the sky. It is embedded in a dusty circumstellar envelope, which was dubbed ``the Fried Egg Nebula'' by \citet{Lagadec2011}, due to its peculiar morphology in mid-IR images (VISIR) showing a double detached shell, with timescales of ejection reaching few hundred years. 

The distance to IRAS~17163, and therefore its class, has been under debate for the past decades. \citet{LeBertre89}, classified it as a proto-planetary nebula (post-AGB star) based on its estimated luminosity ($\sim$10$^{4}$~$\lsun$) and observed spectral characteristics at a distance of 1 kpc. This distance was a rough estimate based on the radial velocity (5.3 km~s$^{-1}$; Local Standard of Rest (LSR)), which was derived using the H$\alpha$ emission, while we should point out that the luminosity was derived based on the observed and not the dereddened photometry. Subsequent studies found this object to be an early A type star, at a distance of 4~kpc, which was estimated using the radial velocities of the interstellar K {\sc i} doublet absorption, yielding a bolometric luminosity of 5$\times$10$^{5}$ \lsun, i.e. a supergiant \citep{Lagadec2011}. This finding confirmed the status of the Fried Egg Nebula as a massive star (estimated mass 25-40 \msun), which is most likely evolving quickly towards and not away from the blue \citep[see e.g.][]{Meynet2015}, and, in all likelihood with a final fate as a supernova. 

IRAS 17163 was suggested by \citet{Lagadec2011} to be one of the very few examples of a star in transition from the Red Supergiant stage to the Wolf-Rayet or Luminous Blue Variable phase (IRC+14020, HD179821, $\rho$~Cas, HR8752). Note that \citet{Smith2004} suggested that YHGs might be missing luminous blue variables (LBVs), also known as S Doradus variables. Such objects change their underlying temperatures and wind properties on timescales of years to decades \citep{Humphreys1994,Vink2002b}. The central star of the Fried Egg Nebula shows P Cygni like H$\alpha$ profiles and slow winds similar to those of YHGs and S Dor variables. A thorough discussion on how to tell those objects apart is presented in \citet{Humphreys2017}. 

\citet{Wallstrom2015} pointed out the difficulty in determining the distance to the star and reported a range of 1-7 kpc depending on the method used. In the current work we use the new parallactic measurements from Gaia \citep[DR2;][Gaia Collaboration]{Gaia2018} and provide a new distance estimation ({\it{Sect.~\ref{Funda}}}), which brings the object as close as 1.2 kpc, and which we adopt in the entire paper. This is three times closer than found in the most recent studies, and therefore a revised analysis on the object is necessary. In this paper we show that despite its closer distance, when the photometry is treated properly accounting for a visual extinction of approximately 10 magnitudes, an effect which was previously neglected, IRAS 17163 maintains its luminosity and therefore its class. 

IRAS 17163 is ideally located in the sky for observations with the Very Large Telescope Interferometer (VLT/VLTI), in order to study its innermost structures and to provide immediate information on this enigmatic evolutionary phase. The dust shells observed in the infrared (IR) are located at large distances from the star ($>$ 1\arcsec) and they are seemingly circular symmetric, but inhomogeneous \citep[e.g. Herschel Space Observatory, VLT/VISIR;][]{Hutsemekers2013,Lagadec2011}. The warm (10~$\mu$m) emission traces the most recent mass-loss episodes, while the origin and the spatial distribution of the 2~$\mu$m emission is being investigated in this study. In the case of the Fried Egg Nebula, the outflowing material tends to become more spherically distributed as the distance from the source grows. Recent observations with the Atacama Compact Array (ACA) of the Fried Egg Nebula revealed the presence of a red-shifted spur ($\sim$20\arcsec), which may trace an unidirectional ejection event \citep{Wallstrom2017}. Investigating the present day mass-loss will provide us with the information on the geometry of the stellar wind. The only other object for which this property was investigated in milli-arcsecond (mas) scales (i.e. present-day wind), is the post-RSG IRC +10420 \citep{Oudmaijer2013}. From NIR interferometric data (AMBER MR and HR) the presence of an ionised, bi-polar flow was inferred, but the sparse uv-coverage made the result model dependent. 

The key evolutionary phase, its brightness and roundness, the sparsity of observed high-mass post-RSGs and its excellent location on the sky have led us to initiate an in-depth, high spatial resolution study of IRAS 17163 and its associated Fried Egg Nebula using spectra and images from VLT and VLTI. 

In {\it{Sect.~\ref{observ}}} we describe our optical and near-infrared spectroscopic (X-Shooter), spectropolarimetric (ISIS/WHT) and interferometric (GRAVITY, AMBER) observations and in {\it{Sect.~\ref{obs_resu}}}, we present the observational results. In {\it{Sect.~\ref{Funda}}} we present the fundamental parameters of IRAS 17163, including the revised distance, its spectral type (i.e. effective temperature), and the dereddened photometry. In {\it{Sect.~\ref{geometry}}} we spatially resolve the inner parts of the nebula, and we trace geometries of the material down to a few au from the central object. To do so we apply simple geometric modeling which allows us to constrain the size and the geometry of the 2~$\mu$m continuum emission, the Br$\gamma$ emission and the Na {\sc i} doublet emission using our VLTI/GRAVITY and AMBER observations. In addition, in the same {\it{Sect.~\ref{geometry}}} we perform and present for the first time the model independent image reconstruction of the object at scales so close to the star (mas scales; few au). In {\it{Sect.~\ref{dust_model}}} we apply a 2D radiative transfer model to simultaneously fit the dereddened available photometry and the radial profiles of VISIR images \citep[8.59 $\mu$m, 11.85 $\mu$m and 12.81 $\mu$m;][]{Lagadec2011} for the first time. Lastly, the discussion and the summary of the paper are presented in {\it{Sect.~\ref{Discussion}}} and {\it{Sect.~\ref{conclusions}}} respectively.


\vspace{-0.02cm}

\section{Observations and data reduction}

\label{observ}

\subsection{Spectroscopy/Spectropolarimetry}

\subsubsection{X-Shooter spectroscopy}

\label{xshooter}

IRAS 17163 ($RA=17^{h}19^{m}49.3^{s}$, $Dec=-39^{o}10\arcmin37.9\arcsec$ [J2000]) was observed during four observing sessions in 2017 using X-Shooter at ESO's VLT. Two short visits were made in April 2017, and two longer ones in May 2017. X-Shooter captures the spectrum of celestial objects from 0.3 to 2.4 $\mu$m in one shot, using three arms which cover the UV-blue (300-560 nm), visible (550-1020 nm) and near-infrared (1020-2480 nm)
parts of the spectrum respectively \citep{odo2006} and their technical overview is presented in Table~\ref{logspec}. The observations were conducted in nodding mode with a nodding step of 6$\arcsec$. The
slit width was 1.0 arcsec in the UV-Blue arm, and 0.4 arcsec in both
the visible and the near-infrared arms. The set-up results in a
spectral resolution of order 10,000 in the VIS and NIR setting and
slightly lower, 9000 in the UVB. 
The data were reduced using the standard ESO pipeline.

Given the extreme colours of IRAS 17163, very faint in the blue ({\it B} $\sim$ 17) moderately bright at {\it V} ($\sim$ 13), while very
bright at near-infrared wavelengths, {\it K} $\sim$ 2.4 \citep[2MASS;][]{Cutri2003}, the observing strategy was adjusted
accordingly.  Observing the blue spectrum required multiple
$\sim$20 minute exposures, while the near-infrared spectra were taken
with short exposure times going down to the minimum possible exposure
times of 0.1 seconds. The total exposure time in the UVB setting is of order 6000s
split over 4 dates. The longest exposures were taken in May 2017.
The VIS setting  was
observed on all days as well, but the main data presented here are
the multiple 60s exposures obtained in May 2017. Most NIR data are
saturated, even those with the shortest exposure times. After visually
inspecting all (hundreds of) exposures, we identified a handful of
spectra that were not affected. These are mostly from the May 29 run.

Considering that this is one of the brightest infrared sources in the
sky, it is interesting to note that the data discussed here are not
only the first high signal-to-noise data in the blue part of the
spectrum, but surprisingly, they constitute also the first
high resolution {\it JHK}-band near-infrared spectra of the object.

\subsubsection{Spectropolarimetry}

The linear spectropolarimetric data were obtained with the ISIS
spectrograph on the WHT, La Palma, during the night of 2015 August
4. Most of the data from that run were published elsewhere
\citep{ababakr17} to which we refer for more details regarding the
observations. Let us briefly summarise these below. A1200R grating
centred at 680 nm, with a spectral range of 100 nm, and a slit width
of 1.0 arcsec providing a spectral resolution of $\sim$35 kms$^{-1}$
was used. The polarisation optics, which consist of a rotating
half-wave plate and a calcite block, were employed in order to perform
linear polarisation observations.
The total integration time was 1800s for IRAS 17163. Polarised
standard stars and zero-polarised standard stars were observed to
calibrate for the instrumental polarisation and angle offset, both of
which were found to be negligible.
 
The data reduction was carried out using {\sc iraf} \citep{Tody93}.
The extracted spectra were imported into the {\sc tsp} package
\citep{1997StaUN66B} to compute the Stokes parameters. Finally,
the percentage of linear polarisation, P, and polarisation angle,
$\theta$, were obtained. The spectrum covers the wavelength range 635
- 720 nm. Due to a combination of the redness of the source and the
response curve of the optics, the signal-to-noise ratio of the
spectrum ranges from $\sim$100 blueward of H$\alpha$ to larger than
300 at redder wavelengths. As the countrates increase very steeply
in the first 10 nm, the polarisation values at these wavelength are
less precise than those longward of around 650 nm.


\subsection{Interferometry}

\subsubsection{GRAVITY observations}

\begin{figure}[h]
\includegraphics[scale=0.35,angle=90]{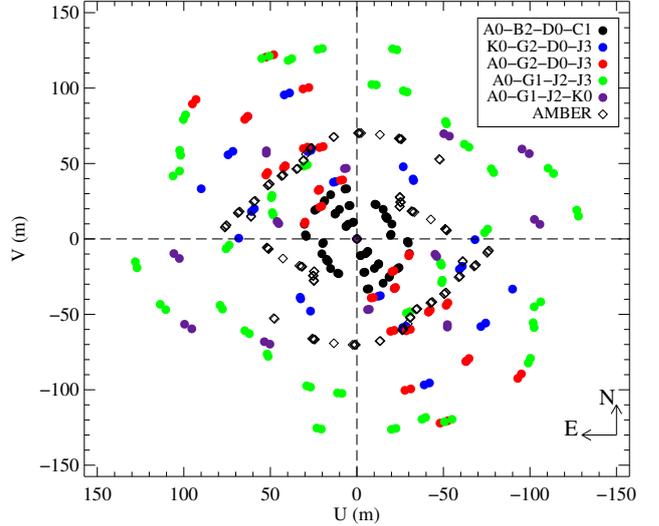}
\caption{uv-plane coverage of VLTI/GRAVITY and AMBER observations of IRAS 17163.}
\label{fig:uvplanefe}
\end{figure}

IRAS 17163 was observed for a total of nine hours during 10 nights between April and August 2017 with GRAVITY \citep{Gravity_Coll2017,Eisenhauer2011} using the four 1.8-m ATs of the VLTI. GRAVITY is an interferometric instrument which operates in the K-band, and it combines signal from the beams of the four telescopes, delivering interferometric observables at a range of spectral resolutions. To achieve an optimal image reconstruction, the data were taken at three different baseline configurations: small (A0--B2--D0--C1), medium (K0--G2--D0--J3, A0--G2--D0--J3) and large (A0--G1--J2--J3, A0--G1--J2--K0) resulting in a good filling of the uv-plane for a 4-element interferometer. The uv-plane coverage is shown in Fig.~\ref{fig:uvplanefe}. 

The observed wavelength range covered the near-IR K-band (1.99 $\mu$m--2.45 $\mu$m) and all the observations were performed in high resolution (HR) combined mode, which provides a spectral resolution of R $\sim$ 4000. This corresponds to a velocity resolution of 75~kms$^{-1}$. The observed configurations combined provided projected baseline lengths B between $\sim$10~m and 130~m, corresponding to a maximum angular resolution of $\lambda/2B\sim 1.7$~mas at the wavelength of Br$\gamma$, that corresponds to a spatial scale of $\sim$2~au at 1.2 kpc. The position angles of the observed baselines covered the entire range between 0-180$^{\circ}$. The technical overview of the observations including the integration times and atmospheric conditions (i.e. coherence time, seeing) is given in Table~\ref{tech}.      

HD~151078 (RA=16$^{h}$46$^{m}$48$^{s}$, Dec=$-$39$^{o}$22\arcmin37.1\arcsec [J2000]), a red giant branch star of K0III spectral type, K magnitude of 3$^{m}$.2 and a uniform disc diameter of 1~mas \citep[JMMC SearchCal;][]{Bonneau2011} was observed as a standard calibration object within 30 min of the science object and under relatively stable weather conditions. HD~151078 was selected as calibrator for all configurations and observing nights. This object was also used as a telluric standard and during the spectra normalisation process.
For the reduction and calibration of the observations the GRAVITY standard pipeline recipes (as provided by ESO, version 1.0.5) were used with their default parameters. Before correcting the telluric absorption features in the spectrum of HD~151078, we removed the absorption features including the CO band in the spectra of the calibration standard object. This could be done by dividing the standard spectra with a telluric corrected spectrum of a standard object of same spectral type (HD~105028) after having it rebinned and having applied a velocity shift to match the resolution and velocity of our observations. The K-band template spectrum of HD105028 was obtained from the NIFS sample of the online library of the GEMINI observatory\footnote{http://www.gemini.edu/sciops/instruments/nearir-resources/spectral-templates/library-v20} and covers a wavelength range of 2.04-2.43 $\mu$m, which is almost the entire range of our spectra.



\subsubsection{AMBER observations}

IRAS 17163 was observed using AMBER in visitor mode on the night of 22nd of June 2012. AMBER is a decommissioned instrument of the VLTI that used to operate in the near-infrared {\it H} and {\it K} bands as spectro-interferometer, combining beams from three telescopes. Our observations were performed using the two baseline configurations D0-I1-H0 and D0-H0-G1, delivering projected baselines between $\sim$30 and $\sim$70 m and position angles PAs between $\sim$5 and $\sim$170$^\circ$ (see Fig.~\ref{fig:uvplanefe} and Table~\ref{amber}). The observations were made using the high spectral resolution mode (R=12000). The obtained spectra cover a wavelength range between 2.15 and 2.19 $\mu$m, which includes the Br$\gamma$ in emission similar to our GRAVITY observations but not the Na {\sc i} and Mg {\sc ii} emission. The technical overview of the observations including the integration times and atmospheric conditions (i.e. coherence time, seeing) is given in Table~\ref{amber}.

The dataset was reduced using the amdlib software \citep[version 3.0.9][]{Tatulli2007,Chelli2009}. During the data reduction we applied frame selection criteria in order to limit the data quality degrading effects that AMBER data generally suffers from. We therefore chose to work with the best 20\% of the observed frames which has been previously suggested to provide consistent visibilities \citep{Malbet2007} and we selected a piston cut-off of 100 $\mu$m since for HR data only large piston values ($>$ 100 $\mu$m) affect the visibility measurements. This process is described in \citet{Tatulli2007}.

Due to the overall adverse weather conditions (seeing $>$ 1\arcsec, coherence time $<$ 5 ms; see Table~\ref{amber}), an absolute calibration could not be performed by only making use of the calibration star as a standard method. Instead, we used our GRAVITY observations of the science object to estimate the emitting size of the continuum at 2 $\mu$m and we used the predicted continuum visibilities of a uniform disk at the corresponding AMBER baselines and PAs to calibrate the AMBER visibilities. We note that the observations are ~5 yr apart, and during this period a typical yellow hypergiant is characterised by significant quasi-period photospheric pulsations, resulting in differences in its radius by up to 50\% \citep[e.g.][]{Lobel1994,deJager1998}. As we show later in the paper (Sect.~\ref{size_e}), the 2 micron emission originates from the stellar object directly, corresponding to the stellar size. Therefore, some degree of uncertainty and bias is to be expected during the calibration process of the AMBER dataset.

 Although the observations are $\sim$5 yr apart, the continuum size at 2 $\mu$m corresponds to the stellar size and is not expected to vary significantly within this period. Therefore, the time difference alone does not introduce significant uncertainties and bias in this calibration process.  

\section{Observational results}

\label{obs_resu}

\subsection{K-band spectrum}

\label{k_band}

\begin{figure*}[h]
\begin{center}  
\includegraphics[scale=0.6]{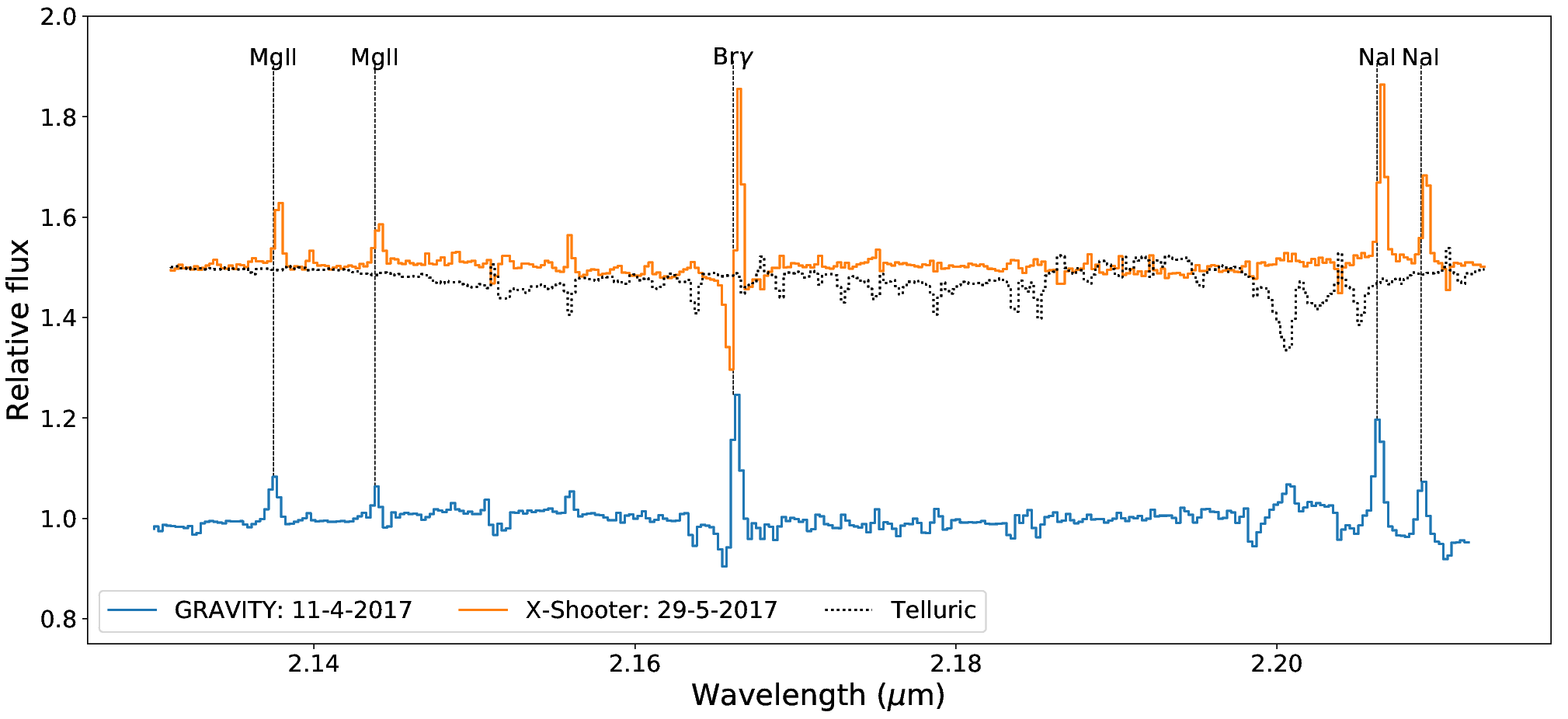} 
\end{center}
\caption{GRAVITY and X-Shooter spectra of IRAS 17163 normalised to the 2~$\mu$m continuum level. The presented wavelength range contains all the observed Br$\gamma$, Na {\sc i} and Mg {\sc ii} emission lines. The dotted line represents the normalised telluric absorption spectrum applied to correct the spectrum of the object. The observed difference in line-to-continuum ratio is due to the higher spectral resolution of the X-Shooter.}
\label{fig:spectrum3}
\end{figure*}

We present a new NIR spectrum of IRAS 17163. Our observed wavelength range includes the emission lines of the Br$\gamma$ hydrogen recombination at 2.167~$\mu$m, the Na {\sc i} 2.206~$\mu$m and 2.209~$\mu$m doublet and the Mg {\sc ii} 2.137~$\mu$m and 2.144~$\mu$m, while there are no signs of CO lines in emission or absorption. The final averaged and corrected spectrum of IRAS 17163 centered on the lines of interest (Br$\gamma$, Na {\sc i} and Mg {\sc ii}) is presented in Figure~\ref{fig:spectrum3}, overplotted with the X-Shooter spectrum. 

\begin{figure}[h]
\begin{center}  
\includegraphics[scale=0.27]{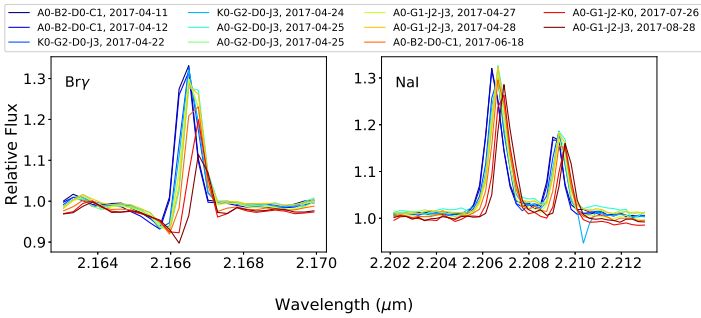} 
\end{center}
\caption{Spectra of the Br$\gamma$ (left) and Na {\sc i} doublet (right), normalised to the continuum level, for the different observing dates and configurations in colours. The dataset shows a gradual drop in the flux of the Br$\gamma$ emission (up to $\sim$67\%) over the observing period of four months, while Na {\sc i} emission does not vary significantly.}
\label{fig:fluxes3}
\end{figure}

Thanks to the GRAVITY observations, we have a time series of 5 months which we can directly compare. Br$\gamma$ shows a P-Cygni profile and its peak intensity gradually drops with time while Na {\sc i} emission shows only emission and does not vary significantly given the uncertainties of $\sim$13\% (Fig.~\ref{fig:fluxes3}). Given that our spectra show clearly a P-Cygni profile we can conclude that Br$\gamma$ is spectrally resolved. In particular, for the first observing date (2017-04-11) the emission part of Br$\gamma$ is characterised by a line to continuum ratio of 1.32 at a peak velocity of 57~kms$^{-1}$ (LSR) and an observed FWHM of $\sim$72~kms$^{-1}$, which is close to the instrumental resolution ($\sim$75~kms$^{-1}$). The dataset shows a gradual drop in the peak of the Br$\gamma$ emission (up to $\sim$67\%) over the observing period of four months. In particular, during the last observing date (2017-08-28), Br$\gamma$ emission is characterised by a line to continuum ratio of 1.11 at a higher peak velocity of 68~kms$^{-1}$ and a FWHM of $\sim$64~kms$^{-1}$. The strongest line of the Na {\sc i} doublet is broader by almost a factor of two ($\sim$130~kms$^{-1}$) compared to Br$\gamma$, with a line to continuum ratio of 1.21 at a peak velocity of 27~kms$^{-1}$. The strongest line among the Mg {\sc ii} shows a FWHM of $\sim$100~kms$^{-1}$ and a line to continuum ratio of 1.13 at a peak velocity of 24~kms$^{-1}$, similarly to Na {\sc i}. The observed lines in X-Shooter spectrum are narrower ($\sim 55$~kms$^{-1}$ for the atomic lines) and have a slightly larger line-to-continuum ratio than in the GRAVITY spectrum, which can be explained by the lower spectral resolution of the GRAVITY data. The line equivalent widths (EW) of the two datasets are similar to within 10\%, the typical error in EW determinations. This suggests that either no extended flux is lost in the GRAVITY observations and that therefore the object is fully within the field of the 1.8m AT (roughly 200~mas), or that any flux loss is the same for the continuum and the lines.
 
\begin{figure}[h]
 \begin{center}  
\includegraphics[scale=0.5]{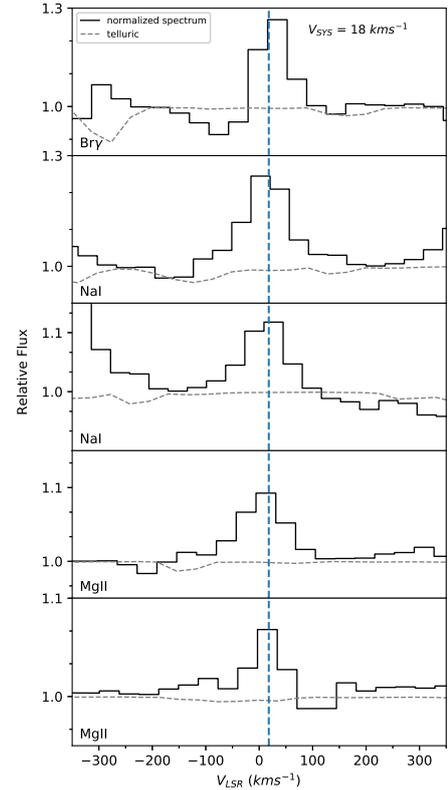} 
\end{center}
\caption{Line profiles of the observed Br$\gamma$, Na {\sc i} and Mg {\sc ii} emission towards IRAS 17163. The dotted line represents the telluric absorption spectrum applied to correct the spectrum of the object. The vertical line represent the systemic velocity as reported by \citet{Wallstrom2015} using the optical Fe {\sc ii} emission.}
\label{fig:spectrum}
\end{figure}

The resulting peak velocities of Na {\sc i} and Mg {\sc ii} are 9~kms$^{-1}$ and 6~kms$^{-1}$ higher than the systemic velocity reported by \citet{Wallstrom2015} (LSR; 18~kms$^{-1}$), who used the Fe {\sc ii} emission lines in the optical. \citet{Wallstrom2017} used the H30$\alpha$ recombination line, but at poorer spectral resolution and found a very similar velocity of 21~$\pm$~3~kms$^{-1}$. The observed difference between our measurements and those of Fe {\sc ii} lines are most likely due to the low spectral resolution of GRAVITY compared to HERMES (High Efficiency and Resolution Mercator Echelle Spectrograph) spectrograph on the 1.2-m Mercator telescope \citep[4~kms$^{-1}$,$\sim$75000;][]{Raskin2011}. The larger observed velocity shift of Br$\gamma$ is most likely due to its P-Cygni profile which results in a blue shift absorption and a peak emission at longer wavelengths. Therefore the comparison of Br$\gamma$ peak velocities with the ones derived from the other lines has little meaning. The spectrum of Br$\gamma$ shows some indications for absorption at longer wavelengths, which needs further investigation. The observed line profiles in velocity space can be seen in Fig.~\ref{fig:spectrum}.

Previous studies have explained spectroscopic variability in P-Cygni H$\alpha$ profiles as a result of variability in mass-loss rate \citep{deGroot1987} accompanied by possible variations in the velocity field \citep{Markova2001}. In particular, a lower mass-loss rate would lead to a more transparent envelope and therefore would give us access to the hotter regions close to the star. With significant variations in mass-loss rates though, one would expect variations in Na {\sc i} emission profiles, which is not what we observe in the time range of our GRAVITY observations (4-5 months). The first and thus far only NIR spectrum of the object was published by \citet{LeBertre89}, at a spectral resolution of R$\sim$1600 in the wavelength window of 2.00-2.40~$\mu$m. As reported and seen in Figure~9 from \citet{LeBertre89}, the only clear emission was Br$\gamma$. A closer look of that spectrum reveals some weak Mg {\sc ii} and Na {\sc i} emission. Mg {\sc ii} may appear weaker than what we see in our spectrum due to the lower spectral resolution. The fact that the ratio of Br$\gamma$ over Na {\sc i} was clearly higher 30 years ago though, and therefore Na {\sc i} emission has become stronger, could indicate the presence of significant variations in mass-loss rates i.e. a recent mass-loss episode. This variability appears to be more significant in a timescale of few decades but not within few months. 

The Na {\sc i} doublet has been previously observed in emission towards similar objects \citep[e.g. IRC+10420;][]{Thompson1977,Oudmaijer2013} while Mg {\sc ii} has not been previously observed towards objects of this class \citep[e.g.;][]{Oudmaijer2013}. \citet{Clark2000} present K-band spectroscopy of Be stars, where they report the presence of Br$\gamma$ and Mg {\sc ii} emission with the absence of HeI and in only few cases also the presence of the Na {\sc i} doublet emission for stars of B2-B4 spectral type. Understanding the origin of the Na {\sc i} doublet emission has been a main challenge in various studies.  In Sect.~\ref{NaI_disc} we investigate several scenarios to explain the observed Na {\sc i} doublet emission. 



\subsection{Continuum polarisation}

\begin{figure}
        \centering
                       \includegraphics[width=8cm,trim={0 100 0 100},clip=true]{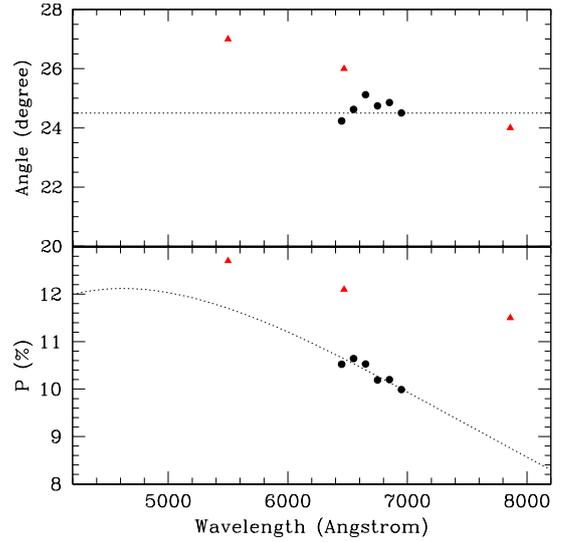}
                       \caption{The polarisation (bottom panel) and
                         polarisation angle (upper panel) observed
                         towards the Fried Egg Nebula. The black dots
                         represent the broadband values, computed in
                         steps of 10 nm, or 100 Angstrom, from the
                         spectropolarimetric data. The red
                         triangles represent the \citet{LeBertre89}
                         data. The dashed line is the best fit
                         Serkowki law for interstellar polarisation.}
\label{cont}
\end{figure}

To arrive at broadband overview of the optical spectro-polarimetry, we binned the
spectrum in steps of 10 nm from 640 nm onwards and show the resulting
polarisation and position angle in Fig.~\ref{cont}. IRAS 17163-3907
shows a very strong continuum polarisation of order $\sim$ 10.5\% at
position angle 25$^{\circ}$. The polarisation declines by almost 1\%
over the spectral range, whereas the position angle varies within 1
degree.

The polarisation can be fitted with a ``Serkowski'' law \citep{1975ApJ196261S}. This describes the empirical wavelength
dependence of the polarisation due to interstellar dust with only 3
parameters, the maximum polarisation, the wavelength at which this
polarisation is maximum, and, $k$, a measure for the shape of
the polarisation behaviour. Keeping the latter at its typical value of
$k$=1.15, we arrive at  our best fit, for $P_{max} = 12.12 \pm 0.4 \%$ at
$\lambda_{max} = 469 \pm 23 $ nm, which is shown in the Fig.~\ref{cont} as a dotted
line. The Serkowski law does represent the observed polarisation very
well. Given the large interstellar extinction towards the object and
the fact the observed polarisation can be described with the Serkowski
law, we can infer that a large fraction of the observed polarisation
is due to foreground material. As pointed out by for example
\citet{oudmaijer01}, 1\% in polarisation corresponds to around 1
magnitude in $A_V$, so the polarisation implies an interstellar $A_V$
of order 10 magnitudes. This is consistent with the value we determine in Sect.~\ref{dered} based on the spectral energy distribution (SED). Note that the large interstellar extinction is supported by both CO observations, which reveal the existence of a foreground dark cloud \citep[e.g.,][]{Dame2001,Wallstrom2015}, and the strong diffuse interstellar bands (DIB) presented in \citet{Lagadec2011}. 


\citet{LeBertre89} also measured a high level of polarisation of
12.7\%, 12.1\% and 11.5\% at angles 27$^{\circ}$, 26$^{\circ}$ and
24$^{\circ}$ in the {\it V, R$_C$, {\rm and} I$_C$} bands. It would
appear that the polarisation at 650 nm (the central wavelength of the
$R_C$ Cousins band) has changed by $\sim$1.5\%, but the angle stayed
constant between 1988 and 2015. In both datasets the polarisation
decreases towards longer wavelengths which is expected from the
Serkowski's law. However, we note that the position angle clearly
changes with wavelength in the 1988 data which cover a large
wavelength range. Our $R$-band data do seem to have the same slope at
the longer wavelengths. This rotation is an indication of an
additional polarising agent at work in addition to the interstellar
polarisation, and this could be potentially be intrinsic to the
source. This notion is re-inforced by the fact that the polarisation
is variable, although, ideally, multi-epoch measurements should be
made with similar equipment to investigate this properly.




%
%

\subsubsection{Line Spectropolarimetry}

\label{spectropolar}

\begin{figure}
        \centering
                               \includegraphics[width=8cm,trim={0 100 0 100},clip=true]{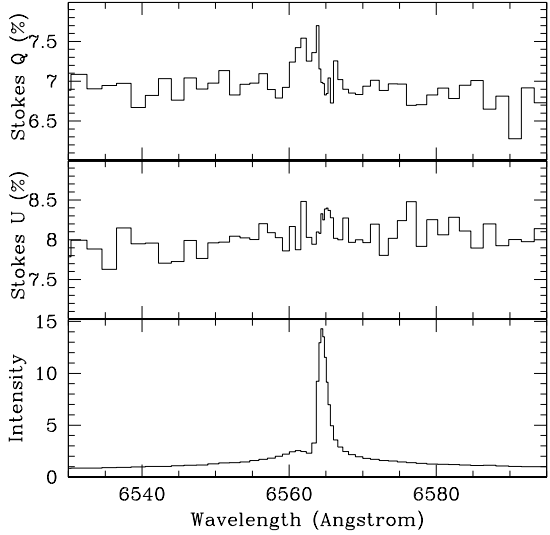}
        \caption{H$\alpha$ spectropolarimetry of IRAS 17163. The
          data are presented as a triplot of the {\it Q, U} and
          intensity Stokes vectors (from top to bottom). The data are
          rebinned to a constant error in polarisation of 0.16\%. The excursion across the $Q$ vector alone indicates an asymmetric scattering geometry at a projected angle of 90$^{\circ}$ across the sky.  }

        \label{iras}
\end{figure}

To examine the polarisation across the H$\alpha$ emission line, we
present in Fig.~\ref{iras} the polarisation spectrum.  It displays the
intensity  and the Stokes $Q, U$ vectors as a
function of wavelength. The data are adaptively binned such that the
statistical error in every wavelength step is the same.  Normally, we
would plot the polarisation and polarisation angle as function of
wavelength
\footnote{We recall that $ P = \sqrt[]{Q^{2}+U^{2}}, $ and $ \theta =
  \frac{1}{2} \arctan \bigg(\frac{U}{Q}\bigg)$.}, but in this
particular situation, the polarisation changes across the H$\alpha$ line are
substantially more clearly defined when the Stokes {\it Q, U} vectors
are displayed. This is because the wavelength independent (across H$\alpha$) interstellar polarisation adds a constant scalar to the {\it Q, U} values, such that they retain their original appearance across H$\alpha$, whereas the polarimetric line profile will be more affected (see equation in footnote; or e.g. Figs 2 and 7 in \citet{ababakr17} and \citet{Oudmaijer1998} respectively). 

The continuum values of $Q$ and $U$ change in such a way that the
total polarisation decreases with wavelength, as outlined
above. Stokes $Q$ displays a clear and significant change across the H$\alpha$ emission line, while Stokes $U$ does not show a clear effect; the excursion that may be present in the few pixels around H$\alpha$ reaches at most the 2 sigma level. A change in $Q$ alone would mean a change in polarisation along a 0$^{\rm \circ}$ angle on the sky, which seems to be the case here. Formally the intrinsic polarisation angle must then be computed using the change in Q, and U, and can be derived using $\Theta=0.5~{\rm arctan}(\Delta U/\Delta Q)$, however, the relatively large uncertainty in measuring any change (if at all) in the $U$ vector across H$\alpha$ means that we cannot measure $\Delta U$ with certainty and a large formal errorbar on the angle of 0$^{\rm \circ}$ would result. A much longer exposure would be necessary to arrive at a more accurate value. As we shall see below the crude measure does agree with what we observe in the interferometry.

The fact that the polarisation across H$\alpha$ is different from the
neighbouring continuum, excludes any explanations for the polarisation
that vary smoothly with wavelength, such as scattering by dust.  The
typical explanation for this so-called ``line-effect'' takes into
account that the star is surrounded by free electrons in ionised gas.
The continuum photons emerging from the stellar photosphere are
polarised by these free electrons. The H$\alpha$ emission line photons
emerge from a larger volume and as a consequence they encounter fewer
free electrons resulting in a lower net polarisation.  If the geometry
of the ionised region is circular on the sky, both line and continuum
photons will have zero net polarisation. When the geometry deviates
from spherical symmetry, a depolarisation across the emission line can
be observed. The resulting polarisation angle is then perpendicular to
the major axis of the asymmetric region on the sky. As the polarisation
is largest when the electron densities are highest, the line-effect is
sensitive to scales often of order stellar radii or less. The
technique has been successfully used to probe asymmetrical structures
at small scales around many types of stars, for example classical Be
stars \citep{1974MNRAS167P27C, 1976ApJ206182P}, Herbig Ae/Be
stars \citep{1999MNRAS305166O, Vink2002,ababakr17} and evolved
stars \citep{trammell94,davies05,patel08}.

If electron-scattering is the explanation for the continuum
polarisation and change in polarisation over the H$\alpha$ line, we can
infer the orientation of the asymmetric material - which would be
perpendicular to the intrinsic polarisation angle of 0$^{\rm
  \circ}$. In most of the references above, this asymmetric material reflects the presence of disks or outflows. However, the current spectropolarimetric data does not allow us to distinguish between the two scenarios. We can say that the spectropolarimetry suggests the presence of an electron-scattering region at scales of order stellar radii and an elongation of this material in the East-West direction.

%

\subsection{Visibilities and Phases}
\label{vis_Sec}

\begin{figure*}[ht]
\begin{center}  
\includegraphics[scale=0.6]{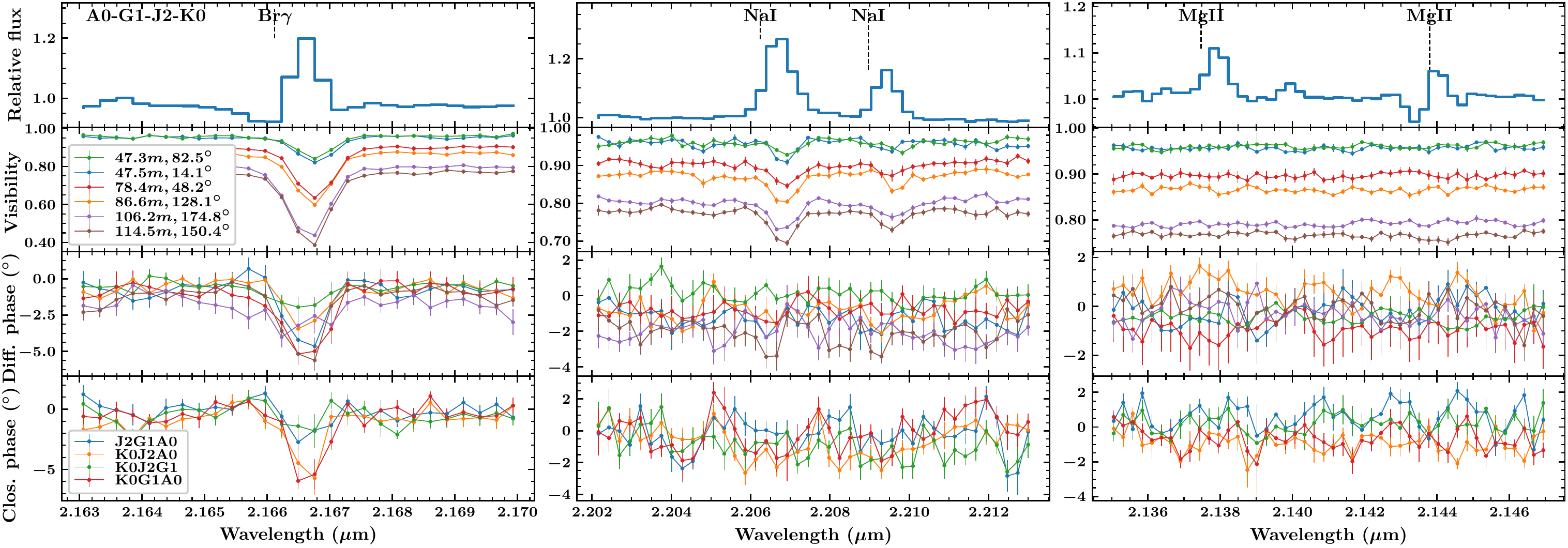} 
\end{center}
\caption{Representative example of relative flux, visibility, differential and closure phase as a function of wavelength around the Br$\gamma$ and Na {\sc i} doublet emission towards IRAS 17163 using GRAVITY on the large configuration. The 6 baselines and 4 triplets of A0-G1-J2-K0 configuration are plotted with different colours. Both Br$\gamma$ emission and Na {\sc i} show a smaller visibility than the continuum while Br$\gamma$ shows a larger drop. Mg {\sc ii} does not show changes in visibility or phases. A differential phase change is also detected towards Br$\gamma$ but not towards Na {\sc i}.}
\label{fig:vis_brg5}
\end{figure*}

\begin{figure*}[ht]
\begin{center}$ 
\begin{array}{cc}
\includegraphics[scale=0.5]{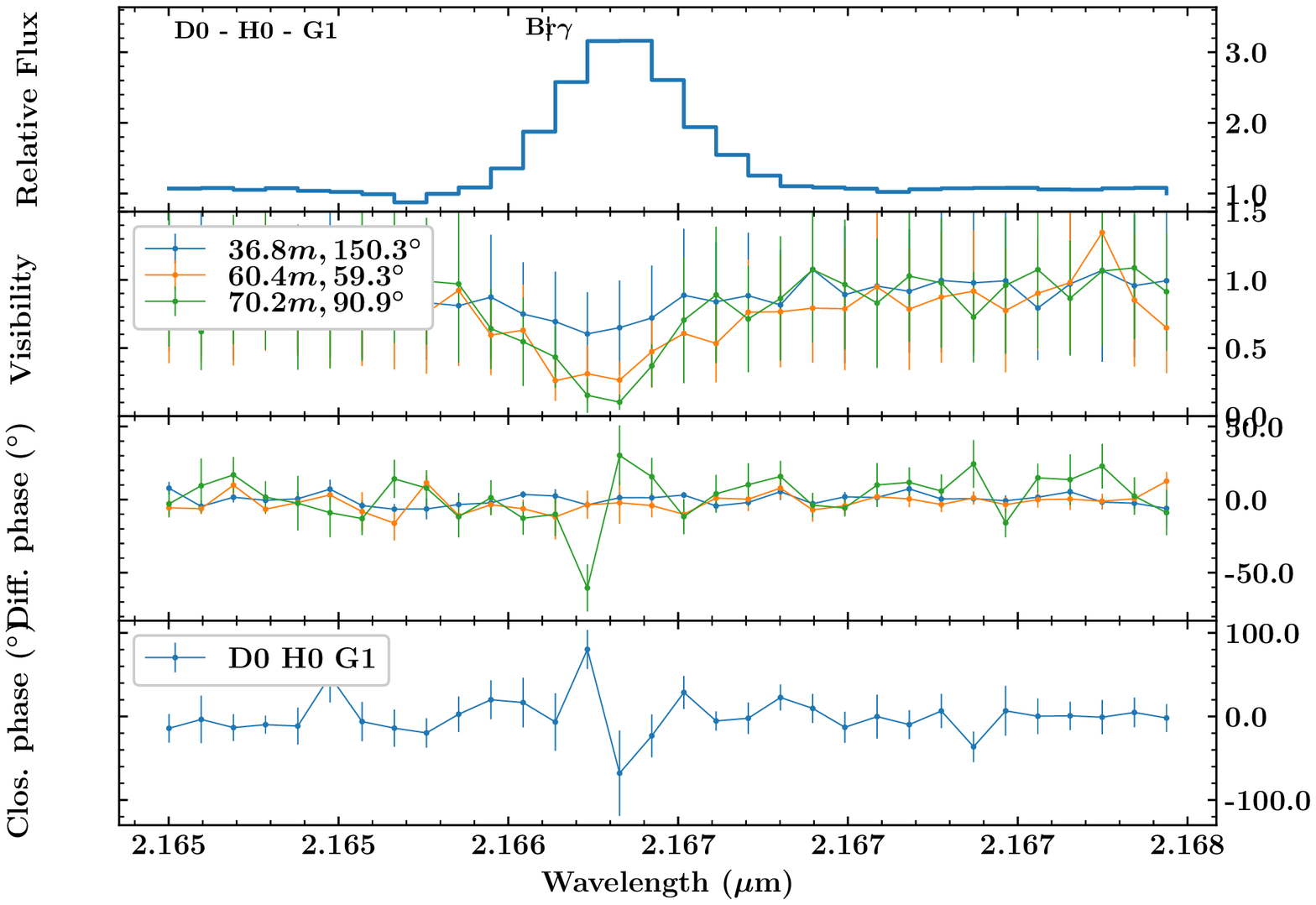} & \includegraphics[scale=0.5]{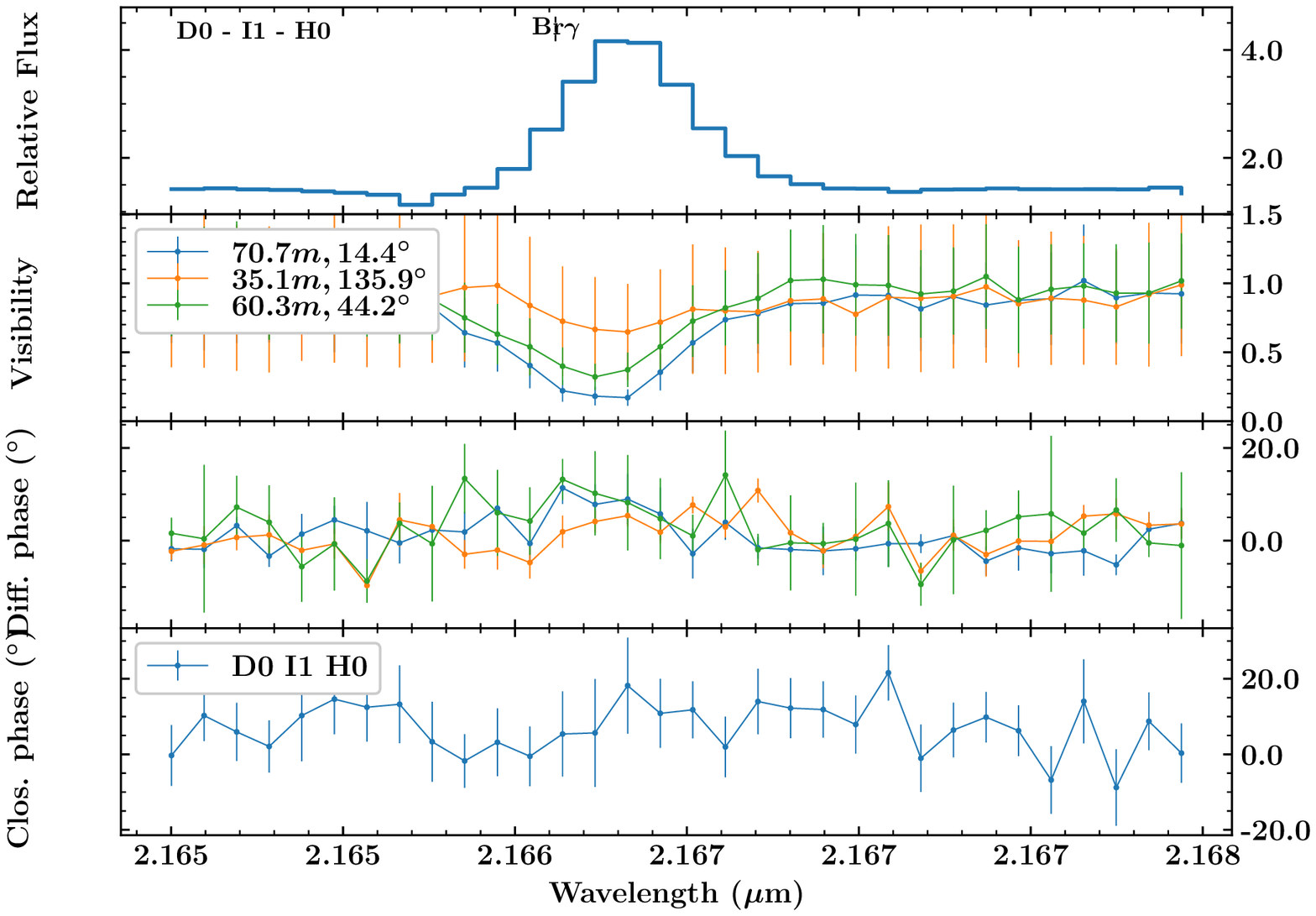} \\
\end{array}$
\end{center}
\caption{Representative examples of relative flux, visibility, differential and closure phase as a function of wavelength for the AMBER D0-H0-G1 and D0-I1-H0 configurations around the Br$\gamma$ emission towards IRAS 17163.}
\label{fig:vis_brg_AMBER}
\end{figure*}

We extracted the visibilities, differential and closure phases from all observed configurations and position angles around the Br$\gamma$ (GRAVITY and AMBER), Na {\sc i} and Mg {\sc ii} (only GRAVITY) towards IRAS 17163. For direct comparison, Figure \ref{fig:vis_brg5} presents the interferometric observables of GRAVITY as a function of baseline length of Br$\gamma$, Na {\sc i} and Mg {\sc ii} emission, only for the large configuration. Similar plots for all configurations can be found in Appendix (\ref{fig:vis_brg1_2}--\ref{fig:vis_brg4}). Figure \ref{fig:vis_brg_AMBER} is similar to Figure \ref{fig:vis_brg5}, but only around the Br$\gamma$, as the wavelength range of our AMBER observations did not cover the Na {\sc i} and Mg {\sc ii} wavelengths. Therefore, our AMBER observations can inform us regarding Br$\gamma$ emission and compare it with the continuum, while the GRAVITY observations inform us also about the additional emission from Na {\sc i} and Mg {\sc ii}. Also note that AMBER combines three telescopes instead of four resulting in a single closure phase. 

Br$\gamma$ shows a different drop in visibility value compared to the continuum for different baselines. In particular it shows deeper drops in visibilities compared to the continuum and the other lines, while it also shows changes in differential and closure phases along the majority of
the baseline position angles. The detected changes are indicative of an emitting region which is more extended
than the continuum emitting region, while the changes in closure phases suggest
that Br$\gamma$ emission stems from an asymmetric region. The line-emitting region (Br$\gamma$) shows a decrease in visibility of 28-35\% for PAs between 80 and 100 degrees (E--W direction) compared to the other observed directions for the longest baseline ($\sim$130~m). This decrease is an order of magnitude larger than the corresponding errors ($\sim$1-5\%, including the uncertainty due to the calibrator and the transfer function) and therefore we attribute the observed difference in geometrical effects, and in particular an elongation of the emission towards E--W direction. No asymmetries in the ionised gas were previously seen/reported towards the object, but this interferometric finding is consistent with our spectropolarimetric results in Sect.~\ref{spectropolar}. 

The observations reveal significant differences among the Br$\gamma$ emission and the line emission from Na {\sc i} and Mg {\sc ii}. 
Na {\sc i} shows a smaller drop in visibilities compared to Br$\gamma$, but no changes in differential
phases and closure phases for the majority of the baselines and position angles. In addition the visibility drop increases as a
function of baseline length for the continuum and the Na {\sc i} emission and independently of PA orientation, indicative of symmetric emission along both axes.
The observations suggest that Na {\sc i} stems from a somewhat larger region compared to the continuum but smaller region compared to Br$\gamma$, and the emission is mostly symmetric at the resolution of the observations. The absence of changes both in visibilities and 
differential phases along Mg {\sc ii} indicate that the line emitting region
follows the distribution of the continuum emitting region, while the absence of
changes in closure phases point towards a symmetric structure of the emitting area. Therefore, we
do not include Mg {\sc ii} in our more detailed analysis assuming that our geometrical results about the continuum emission apply also for Mg {\sc ii}.

\section{Fundamental parameters of IRAS 17163}
\label{Funda}

\subsection{Spectral type}

\label{spectral}

Before proceeding to a detailed modeling of the Fried Egg Nebula, it is important to put constraints on the effective temperature of its central star, IRAS~17163. Thus far, it had not been possible to determine the spectral type in a classical manner for the object; optical spectra up until now were
only usable beyond around 500 nm, where the spectrum is dominated by
emission lines, while it is the properties of the absorption lines
that determine the spectral type. Moreover, spectral classifications are traditionally carried out in the blue spectral region. The blue part of the optical spectrum obtained here is dominated by absorption and can serve
as a basis to determine the spectral type the object, while at the same time
facilitate a comparison with the spectrum of the known post-Red
Supergiant IRC+10420, whose emission spectrum had been shown by
\citet{Wallstrom2015} to be very similar to that of IRAS 17163 at longer wavelengths. Lastly, the 2017 spectrum should also serve as a reference for future studies into any temperature evolution.

\begin{figure*}
        \centering
        \includegraphics[width=0.75\textwidth,trim={0 100 0 100},clip=true]{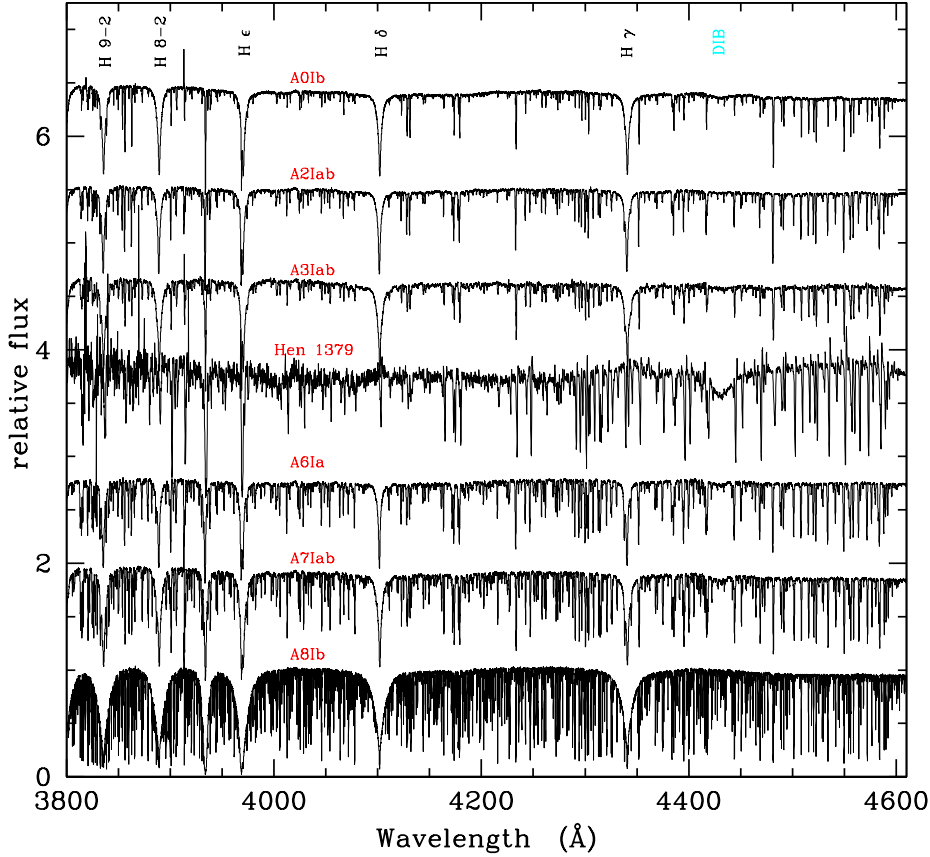}
        \caption{Spectra of IRAS 17163 with a number of spectral
          standards taken from the UVES spectral archive
          (\citealt{bagnulo03}). The spectra are normalised to a continuum value of 1 and offset, for clarity, in multiples of 0.9.}
    \label{spstan}
\end{figure*}

In order to carry out the spectral type classification, we downloaded high
spectral resolution data of a number of supergiants from the
\citet{bagnulo03} spectral database. The spectrum of IRAS 17163 and
that of a number of A-type supergiants is presented in
Fig.~\ref{spstan} in the typical spectral type classification wavelength range,
380-460 nm. Apart from the strong DIB at
$\sim$440 nm betraying a large interstellar extinction, emission filling in the hydrogen absorption lines, and the lines longward of 440 nm affected by emission, the spectrum fits in
naturally with the standard-type supergiants. Indeed, it can be seen
that the object mostly follows the clear progression of the spectral
types when placed between the A6 and A3-type supergiants with a
preference for a slightly earlier than A6Ia spectral type. With
temperatures for A3I and A5I supergiants listed by
\citet{straizys81} of 8892 K and 8298 K respectively, we would infer a
temperature of about 8300-8500 K for the central star of the Fried Egg Nebula.

\begin{figure*}
        \centering
                \includegraphics[width=0.95\textwidth,trim={0 120 0 80},clip=true]{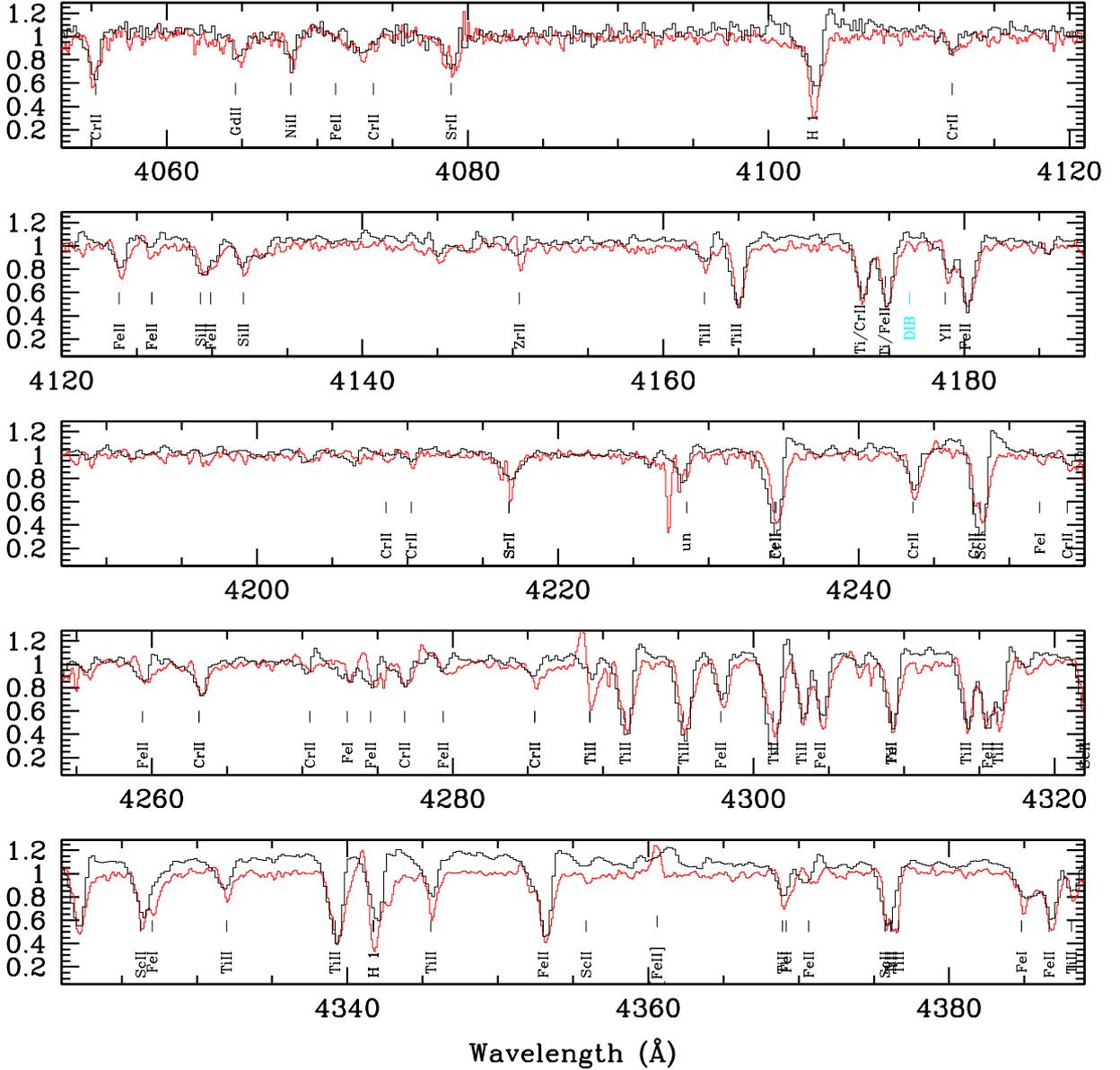}
        \caption{Overlay of part of the blue spectrum  of IRAS 17163 (black line) and IRC +10420 (red line). Note the tantalising similarity, as previously noticed also by \citet{Wallstrom2015} for the optical spectrum above 5000 $\AA$ which is dominated by emission lines.}
    \label{spstanirc}
\end{figure*}

A further clue to the temperature of IRAS 17163 can
be gleaned from the comparison with the well-known post-Red Supergiant
IRC +10420. Figure~\ref{spstanirc} shows a part of the blue absorption
spectra of both objects (the IRC +10420 spectrum taken from
\citealt{oudmaijer98}). It can be seen that the stars are remarkably
similar. \citet{klochkova97} determined a temperature of $8500 \pm 250$~K for IRC+10420, based on intensity of Fe {\sc i} absorption lines. In turn we can infer a similar temperature for IRAS 17163 as well. In the following sections we will adopt a temperature of 8500 K for the central star of the Fried Egg Nebula.


\subsection{Distance}

\label{dist}
The distance of IRAS~17163, has been a debated topic for the past decades (see also Sect.~\ref{intro}), but its accurate determination is crucial in order to constrain the fundamental parameters of both the stellar object (size, luminosity, class) and its surrounding shells (size, mass-loss rates, kinematic time-scales). In this work we take advantage of the new Gaia parallactic measurements \citep[DR2;][Gaia Collaboration]{Gaia2018} and derive a distance of d=1.2$^{+0.4}_{-0.2}$~kpc, which we adopt in the entire paper. To do so we applied the method proposed by \citet{BailerJones2018}, which uses a Galactic Model to weakly inform a distance prior to be used in the determination of the distances, and not just the inversion of the parallax (0.83$\pm$0.17~mas), as the latter is found to generally introduce more biases, failing to treat properly nonlinear effects. For the calculation we made use of the ARI's Gaia services\footnote{http://gaia.ari.uni-heidelberg.de/tap.html}. 

Given that the specific assumptions of this method do not necessarily apply to this object, we also determined the distance using the direct parallax inversion, and found it to be 1.2 kpc, which is in perfect agreement with the one using the method suggested by \citet{BailerJones2018}.  

Gaia parallax presumes that the measured star is a single and point-like source. Our analysis in Sect.~\ref{size_e} suggests that the K band continuum and Hydrogen line emission are confined to the scale of 1 to 1.9 mas, and thus point-like with respect to the Gaia resolution (0.1\arcsec). It also appears that most/all of the K band light is contained within the field of view. Thus the resolved sizes actually support a point-like appearance for Gaia which gives us a further confidence in using the parallactic distance estimation.

\subsection{Dereddened Photometry}

\label{dered}

\begin{figure}[ht]
\begin{center}  
\includegraphics[scale=0.55]{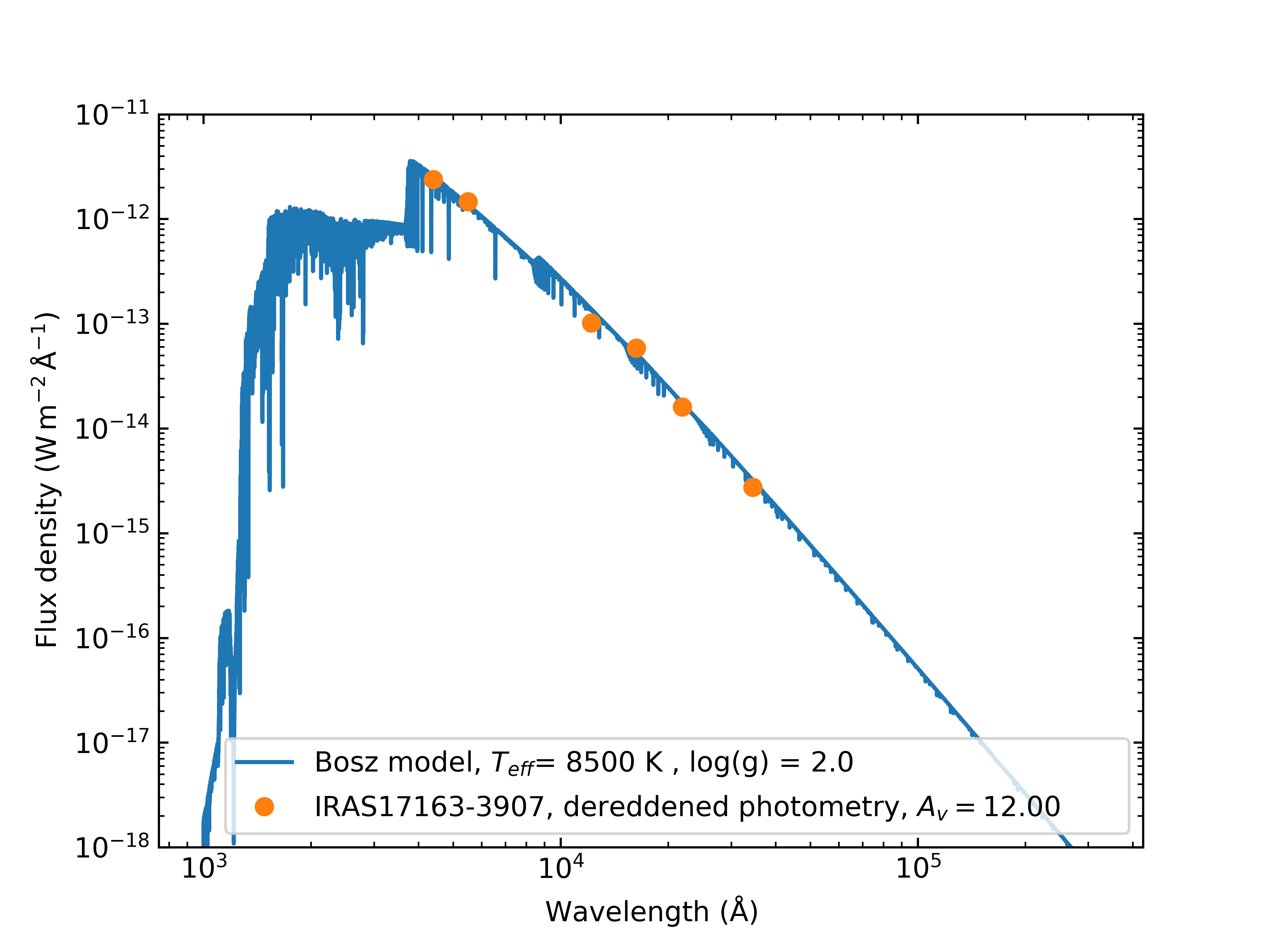} 
\end{center}
\caption{Dereddened IRAS 17163 photometry with total extinction $A_{V}$ = 12.0 to fit a $T_{\rm eff}$ = 8500K Bosz stellar model.} 
\label{photfit}
\end{figure}

The photometry of IRAS 17163 shows strong excess emission beyond 5~$\mu$m. A black body emission assuming a temperature of 8500 K (see Sect.~\ref{spectral}) can reproduce the observed dereddened photometry at wavelengths $<$ 5~$\mu$m, for which the observed flux is solely due to the central heating source. We fit a Bosz stellar model \citep{Meszaros2012} to the photometry of 2.2~$\mu$m and below \citep[including bands: B, V, J, H, K and L;][]{Epchtein87,LeBertre89,Cutri2003}. For the Bosz model we adopted an effective stellar temperature $T_{\rm eff,\star}$ of 8500 K, a surface gravity $\log$~$g$ of 2, which is the lowest available value in the models for the adopted temperature, and a solar composition (metallicity [M/H], carbon and alpha-element abundances). This temperature is an upper limit given the spectral type of IRAS 17163 and the interferometric continuum size (Sect.~\ref{spectral}). The observed photometry is affected by extinction even at shorter wavelengths and therefore needs to be dereddened. We used a least squares minimisation method and found a total extinction of $A_{V}$ = 12.0$^{+0.8}_{-0.33}$ (Figure~\ref{photfit}) after applying the selective visual extinction $R_{V}$-dependent law from \citet{Cardelli1989} for $R_{V}$ = 3.1 which corresponds to a galactic ISM dust \citep{Fitzpatrick2009}, and zero point magnitude from \citet{Bessell1979}. The observed and dereddened photometry is presented in Table~\ref{photometry_data}.  

Our dereddened photometry, when compared to that by \citet{Hutsemekers2013}, shows that the integrated energy flux per unit wavelength of the star in that study is smaller by a factor of $\sim$10 at a distance of 4~kpc, and therefore either the luminosity of the object is $\sim$10 higher than previously thought, or its distance is $\sim$3 times closer. This was an oversight that was propagated in more studies, since both \citet{Lagadec2011} and \citet{Hutsemekers2013} used the photometry presented in \citet{LeBertre89} as a reference, who had not corrected for reddening. In conclusion, based also on Sect.~\ref{de_sed}, integrating the best fit SED after treating properly the observed photometry (i.e. dereddening) in combination with the closer distance of IRAS~17163 at 1.2 kpc, results in a luminosity of $L=5\times10^{5}~\lsun$, and therefore IRAS~17163 maintains its class as a yellow hypergiant. Given its luminosity and effective temperature, we estimate its size to be $R=325~\rsun$. Assuming a lower mass limit of $M=25~\msun$ \citep[$M=25-40~\msun$;][]{Lagadec2011}, the resulting surface gravity $\log$~$g$ is $\sim$ 0.9, which is a typical value for a yellow hypergiant. Although this is lower than the adopted value of $\log$~$g$ = 2, during our fitting process we found that for a given effective temperature, different values of surface gravity did not significantly affect the colours. Therefore the adopted $\log$~$g$ does not affect the reliability of the derived extinction within its uncertainties.

\section{Geometry of the emission}
\label{geometry}

\subsection{Size estimations}

\label{size_e}

The present GRAVITY dataset includes long baselines up to $\sim$130~m and it can be used to estimate the sizes of the continuum emission, and those of the neutral and ionised gas as traced from the Na {\sc i} and Br$\gamma$ emission respectively. For this purpose, we use the calibrated visibilities as measured at the spectral channel of the peak of the emitting lines and the calibrated visibilities of the continuum. To derive and model the actual visibility of the lines (Na {\sc i} and Br$\gamma$), deprived from the continuum contributions, we used Equation \ref{contin}, which describes the total visibility ($V_{line+cont}$) in terms of continuum and line visibilities ($V$) and fluxes ($F$) when dealing with multi-component sources, and solving for $V_{line}$:

\begin{equation}
V_{line+cont} = \frac{V_{cont} \times F_{cont} + V_{line} \times F_{line}}{F_{cont} + F_{line}}
\label{contin}
\end{equation}

\begin{figure}[h]
 \includegraphics[scale=0.35,angle=90]{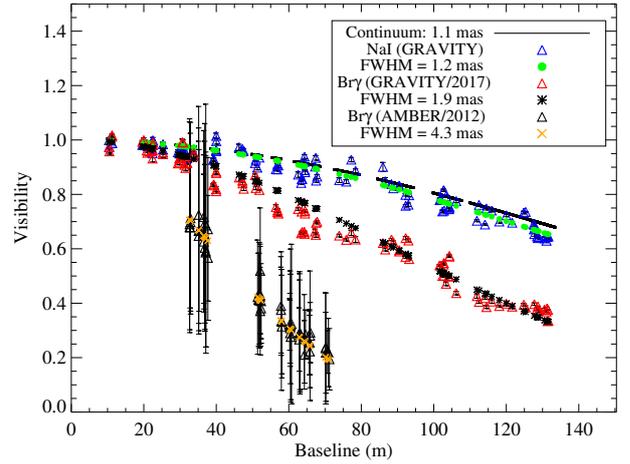}
\caption{Visibilities at the peak of the Br$\gamma$, Na {\sc i} emission lines and continuum emission towards IRAS 17163 overplotted with simple geometrical Gaussian models. Interestingly, Na {\sc i} and continuum emission follow very similar distribution. Na {\sc i} shows smaller drop in visibilities compared to Br$\gamma$, indicating that Na {\sc i} stems from a smaller emitting area compared to the Br$\gamma$. The AMBER dataset was rebinned to match the spectral resolution of GRAVITY. A period of 5 yr indicates a shrinkage of the Br$\gamma$ emitting region to half size.}
\label{fig:Na IBrg2}
\end{figure}

For consistent comparison of the sizes of the line and continuum emission, we fit the observables with simple geometrical models of a Gaussian-shaped emitting region for a range of sizes (0.2~mas--5~mas). We find that the best fit results in a full-width-at-half maximum (FWHM) of 1.09$\pm$0.01~mas for the continuum, 1.20$\pm$0.01~mas for the Na {\sc i} and 1.90$\pm$0.01~mas for Br$\gamma$. Figure~\ref{fig:Na IBrg2} shows the observed values overplotted with the corresponding best fit models for the three cases.  

Taking into account the extreme upper temperature limit of T$_{dust}$ $\sim$ 2000~K \citep[pyroxene;][]{Kobayashi2011} at which dust sublimation occurs, and the expected radius and temperature of IRAS~17163, we find that dust cannot survive at distances $<$ 7~mas ($d = R_\star T_\star^{2}$ $/$ $2 T_d^{2}$) which is larger than the size of Na {\sc i} and continuum emission ($\sim$ 1~mas) by a factor of 7. We point out that dust sublimation of crystalline and oxygen-rich dust occurs at lower temperatures \citep[$\sim$1200~K;][]{Blasius2012}, which is a more realistic approach to IRAS 17163, and therefore 7~mas is only a lower limit for the dust sublimation radius. We conclude that the continuum emission at these wavelengths (2.0 -- 2.4 $\mu$m) originates from the stellar object directly. This finding is also supported by the photometry findings at shorter wavelengths in Sect.~\ref{dered}, where a black body could fit the photometry at wavelengths $<$ 5 $\mu$m. 

Based on our findings in Sect.~\ref{dered} the expected size of the central star is 2.5~mas (diameter; $\sim$650~\rsun) at the distance of 1.2~kpc. The estimated size of the continuum is 1.09$\pm$0.01~mas, and therefore at first there is notable discrepancy between the calculated size and the measured one. Given the geometrical complexity of stellar atmospheres of evolved stars, this difference is most likely due to our simple model of Gaussian distribution, which was selected for a consistent comparison with the line emissions. When one adopts a uniform disk to model the interferometric observables, the continuum size is 1.73$\pm$0.02~mas, which is closer to the expected. The intensity profile that characterises such objects is not a uniform disk nor a Gaussian distribution, and it is strongly characterised by limb-darkening signatures, and even more complex structures are predicted by theory \citep[e.g. convective patterns][]{Chiavassa2011}. 

Having inferred the effective temperature in Sect.~\ref{spectral}, a robust determination of the physical size of the atmosphere would lead to a more accurate estimation of the luminosity of the star. This is something we could not achieve with our current geometrical models, but could be achieved with the use of more advanced atmospheric models, similar to what has been done for Red Supergiants in the past \citep[e.g.][]{Paumard2014}, which are currently lacking for YHGs and it is beyond the scope of this work. For a more robust comparison, one should also consider the uncertainties in the luminosity of the source and the Gaia distance. In addition, we should note that the reported uncertainties in the size estimations represent the goodness of fit and are rather underestimated. In conclusion, considering all the aforementioned uncertainties, the measured angular diameter is in reasonable agreement with the photometric one.

\subsection{Model independent image reconstruction}

\label{recon}

The VLTI is a very powerful interferometer providing high angular resolution in the visible and near--infrared.
The measurements coming from the observations are not direct images though and therefore using image
reconstruction algorithms is crucial if one wants to make a full use of its imaging capabilities.  

We perform and present for the first time the image reconstruction of IRAS 17163 and its close vicinity at mas resolution. To do so we used the MIRA software \citep[Multi-aperture Image Reconstruction Algorithm;][]{Thibaut2008}. During this process MIRA fits the observed squared visibilities (power spectrum) and closure phases (bispectrum) with models, using some a priori image properties (positivity, normalisation and regularisation) and a cost--estimate optimisation. For the cost--estimate optimisation we used the L2-L1 regularisation which assumes the pixel positivity and the compactness of the source. With only 4 telescopes the above constraints are necessary to properly interpolate the ``gaps'' in the coverage of the uv-plane. The maximum number of evaluations of the cost function was set to 200 with a global weight of regularisation of 10$^{-10}$. The regularisation type was set to entropy. In our approach an exact Fourier transforms algorithm was set. For a detailed description of the chosen parameters see \citet{Thibaut2008}.

The computed field of view was $\sim$75~mas for B$_{min}$ = 11 m (2.44~$\lambda$/B$_{min}$; 4~$\times$ $\Theta_{ max}$ to ensure we do not cut out emission), where B$_{min}$ is the shortest projected baseline, and $\lambda$ = 2.2 $\mu$m. Although previous studies have shown that the image reconstruction result did not significantly depend on the required starting image \citep{Millour2009}, we chose to work with multiple simple randomised start images of the same dimensions that ``host'' a point or more extended source at the center. We ran the algorithm for a range of pixel sizes (0.1--1~mas) and we found that for a pixel size $>$ 0.5~mas the resulting image does not change significantly while for smaller pixels (0.1--0.5~mas) finer structures could be seen. To avoid bias introduced by the particular image model (e.g. pixel shape), the spatial resolution of the model should be well beyond the resolution limit of the observations ($<$ 1.7~mas). As a general approach, a super resolution is usually used and the pixel size is determined by $\Delta\theta < \lambda / 4 B_{max}$ ($\sim$ 0.8~mas). To avoid the introduction of artifacts a pixel size of 0.5~mas was chosen. A bandwidth of 4$\times$10$^{-4}$~$\mu$m was set for the Br$\gamma$ and Na {\sc i} image reconstruction corresponding to the width of one spectral channel, while the bandwidth of the continuum was set to two orders of magnitude higher (0.03~$\mu$m). 

\begin{figure}[ht]
\begin{center}  
\includegraphics[scale=0.2]{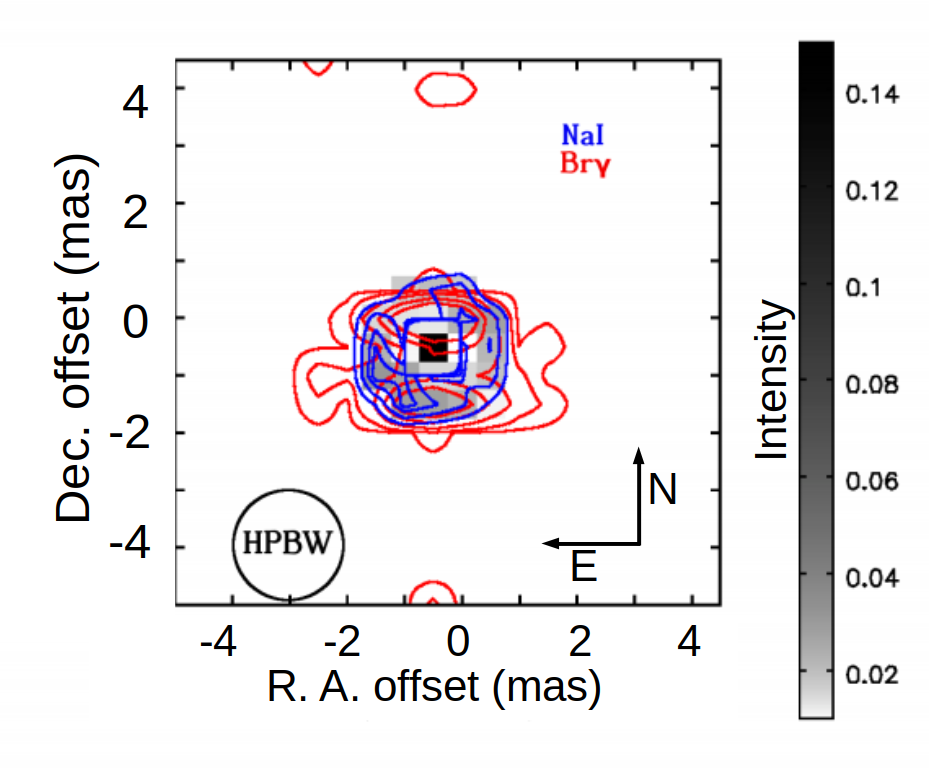} 
\end{center}
\caption{Image reconstruction of the continuum emission (grayscale) overplotted with Br$\gamma$ in red contours at [0.1, 0.3, 0.6, 0.75, 0.95]$\times$peak and Na {\sc i} in blue contours at [0.2, 0.5, 0.8, 0.95]$\times$peak towards IRAS 17163. Br$\gamma$ covers a more extended region compared to the continuum and Na {\sc i}, which appear to follow very similar distributions.}
\label{fig:imageBrgcont}
\end{figure}

Separate image reconstruction was performed for the continuum and around the central channel of the Br$\gamma$ and Na {\sc i} emission (Figure~\ref{fig:imageBrgcont}). The image reconstruction reveals a compact symmetric emission of the continuum and Na {\sc i}, while a more extended Br$\gamma$ emission along with asymmetric features, confirming what we already discussed in Sect.~\ref{vis_Sec}. In particular the image reconstruction for Br$\gamma$ also reveals a northern and southern component.

In conclusion we present a model independent image reconstruction of this object at 2 $\mu$m for the first time. Although we treated this delicate procedure with extra care, the observed flux changes of the Br$\gamma$ at different epochs introduce more uncertainty in the reconstructed image by possibly affecting the observed visibilities. The channel maps of the Br$\gamma$ emission together with the combined (AMBER and GRAVITY) image reconstruction, using three different software packages is less robust and is presented in Appendix~\ref{channel},\ref{bsmem}. Furthermore, Fig.~\ref{fig:Na IBrg2} shows that the object is formally only partially resolved (i.e. does not reach the first zero of visibility), and therefore at the smallest observed scales the image suffers from higher uncertainty. Despite the uncertainties introduced by the special features of this unique dataset, we find that the main spatial structures confirm our findings when applying geometrical models to the interferometric observables, showing a more extended and asymmetric Br$\gamma$ emission than that of the continuum and sodium.   



\section{Modeling the shells of the Fried Egg Nebula}
\label{dust_model}

\subsection{Modeling the warm dust emission}

In order to get quantitative information on the recent mass-loss history as traced by the dust and derive the dust mass and the morphology of the distinct dust shells surrounding IRAS 17163, one has to model the warm dust emission in various wavelengths. In previous works, both \citet{Lagadec2011} and \citet{Hutsemekers2013} assumed a distance of 4~kpc and fit the available photometry with radiative transfer models to estimate the mass within their observed detached shells. \citet{Lagadec2011} focused on fitting a small volume encompassing two warm shells observed in mid infrared wavelengths using VISIR (the ``Fried Egg Nebula'') with a radius of $\sim$ 0.45\arcsec to 2.5\arcsec. \citet{Hutsemekers2013}, on the other hand, focused on fitting a larger cooler volume encompassing of both the Fried Egg Nebula as an unresolved single shell, and a much larger shell revealed at the far-infrared, 70 $\mu$m and 160 $\mu$m images using PACS/Herschel with a radius of $\sim$ 18\arcsec~to 40\arcsec. We note that although both studies achieved a good fit to the SED (apart from \citet{Hutsemekers2013} to the optical part), the resulting total dust mass for the Fried Egg Nebula differs by an order of magnitude. 

The degeneracy can be broken if one not only fits the SED (i.e. constrain the optical depth of each shell) but the intensity profiles as well, which will constrain the position and thickness of the separate shells and the dust temperature. In this study we adopt a model to simultaneously fit a) the dereddened photometry with a newly determined value of visual extinction based on our revised distance and b) the VISIR/VLT infrared radial profiles of the star and the surrounding ``Fried Egg Nebula'', breaking the SED degeneracy.

\subsubsection{Dust modeling: Initialising 2-Dust}
\label{2dust}

In this work we used the 2D radiative transfer code, 2-Dust \citep{Ueta2003}, towards the Fried Egg Nebula to simultaneously fit for the first time both the SED and the flux maps (radial profiles) of chosen wavelengths (8.59 $\mu$m, 11.85 $\mu$m, 12.81 $\mu$m), which allows us to spatially constrain the observed shells (R$_{in}$, R$_{out}$) and their properties (i.e. opacity, temperature, mass). The selected wavelengths are based on the VISIR images of the object, published in \citet{Lagadec2011}. 

2-Dust solves the equation of radiative transfer tracing the heating mechanisms of a grain located at any place in the shell radiated from both the central star and from surrounding dust radiation using a 2D polar grid. The code deals with a large range of input parameters including the fundamental parameters of the central heating object (i.e. effective temperature, size, distance) and the properties of the dust (Sect.~\ref{comp_size}) that surrounds it (i.e. shell size, dust composition, opacities). 

In particular, the stellar parameters and the distance of IRAS~17163 were determined in Sect.~\ref{Funda}, and therefore were fixed to $T_{\rm eff}$ = 8500~K, $R=325~\rsun$ ($L=5\times10^{5}~\lsun$) and $d = 1.2$~kpc during the entire fitting process (IRAS~17163; Table~\ref{2dustResults}). In our approach we refined a polar grid of 48 radial and 8 latitudinal grid points. The log-spaced radial grids for the shells are characterised by an inner and an outer shell radius (R$_{in}$, R$_{out}$). The locations and sizes of the dusty shells were adjusted accordingly to fit the observed radial profiles. In particular, the images and radial profiles of the Fried Egg Nebula at the chosen wavelengths, have previously revealed two shells with further indications of a third hot shell which had not been accounted for during the modeling process \citep{Lagadec2011}. In this work we provide evidence for the existence of a third hot inner shell which we model, resulting in a better fit to both the SED and the radial profiles. In particular, we model three shells, a hot inner shell extending between 0.3\arcsec and 0.45\arcsec, an intermediate warm shell extending between 0.6\arcsec and 1.1\arcsec and an outer cooler shell between 1.9\arcsec and 2.5\arcsec, which we implemented in our model during the SED and radial profile fitting process (Figure~\ref{visir_images_fit}). The ``outermost'' cooler shell reported in \citet{Hutsemekers2013} of a radius of $\sim$ 18\arcsec~to 40\arcsec was not included in our models. The r$_{d,min}$ of the intermediate and the outer shells was strongly constrained at 0.6\arcsec and 1.9\arcsec~respectively by the observed peak of each of the shells within the intensity profile using the existing radial profiles of the images at mid infrared (Figure~\ref{visir_images_fit}). Constraining the r$_{d,max}$ of each of the shells was less robust. The exact position and size of the shells are both important in determining the temperature and the 10 $\mu$m to 18 $\mu$m flux ratio of each shell. 

To be able to model the three distinct shells with 2-Dust, one has to modify the density distribution by applying different values of optical depths $\tau_\lambda$, at each of the individual shells. To do so, we treated each shell separately, using the output flux distribution (flux density over wavelength) of the central star and the inner hot shell as the input for the intermediate shell and so forth. The optical depth $\tau$ applied to each shell, was the only free parameter in our models and was fine tuned to provide a good fit to the relative peak intensity profiles of the observed 8.59, 11.85 and 12.81 $\mu$m emission, in combination with a good SED fit. Details on the fitting and the final results are presented in Sect.~\ref{de_sed}. 


\subsubsection{Dust composition and size}

\label{comp_size}

IRAS 17163 is dominated by silicates, which were first observed in the IRAS spectrum of the source as strong emission around 10~$\mu$m and 18~$\mu$m \citep{LeBertre89}, and including pyroxenes (Mg$_{x}$Fe$_{1-x}$SiO$_{3}$). This is indicative of an oxygen-rich composition and it is similar to what is also found towards IRC+10420, Wray 15-751 and other YHGs \citep{Oudmaijer1996,Voors2000}. In this work we adopted a dust composition of a 50/50 mixture of Mg(40)/Fe(60) and olivines \citep{Dorschner1995} with astronomical silicates \citep{Weingartner1999}, similar to those presented in \citet{Hutsemekers2013}. The dust grain size was determined using the MRN size distribution n($\alpha$) $\propto$ $\alpha^{-3.5}$ with $\alpha_{min}$ $<$ $\alpha$ $<$ $\alpha_{max}$ and $\alpha_{min}$ $=$ 10$^{-3}$~$\mu$m. The dust grains that surround IRAS 17163 can have a maximum size $\alpha_{max}$ between 1 and 3~$\mu$m \citep{Hutsemekers2013}. The bulk density is fixed to 3.2 g~cm$^{-3}$. Adopting a maximum grain size of 1 or 3 $\mu$m affects the absorption and scattering cross sections. In particular, for wavelengths $>$ few microns, the scattering cross sections are found to be $>$ 1 order of magnitude larger for an upper limit on the grain size of 3 $\mu$m compared to 1 $\mu$m. The silicate features at 10 $\mu$m and 18 $\mu$m can be seen in both cases. Smaller grains but with constant bulk density, result in lower mass grains and therefore a larger amount of grains within one shell, which increases the total flux within the shell. 


Therefore, the absorption and scattering cross sections generated by the code are dependent on the dust composition and the grain size distribution, altogether influencing the shapes of the resulting SED and radial profiles for each wavelength. In our case a maximum grain size of $\alpha_{max}$ $=$ 1 $\mu$m appeared to provide a better fit of the silicate features in the SED and the intensity profiles compared to the $\alpha_{max}$ $=$ 3 $\mu$m. Adopting different dust types and mixtures is expected to affect the modelled emission but such an approach is beyond the scope of the current work.

\subsubsection{Best fit: SED and synthetic radial profiles}
\label{de_sed}

During the fitting process, the optical depth $\tau$ of each shell, which influences the strength of the emission, was tuned with the goal to reproduce a good fit on the relative peak intensity profiles of the observed 8.59, 11.85 and 12.81 $\mu$m emission. We started the initial fitting process by trying to reproduce the resulting mass-loss rates reported in \citet{Hutsemekers2013}. Then we scaled everything according to the new distance, which meant that the optical depths had to be reduced by a factor of $\sim$ 10 in order to reproduce the updated flux of the shells. We expanded our approach to three shells and varied the values of the optical depths around a central value (up to a factor of 2), which helped us to constrain the solution that appeared to provide a good fit to the observations. 

Our best fit was achieved for a $\tau_{70\mu m}$ of 3.5$\times$ $10^{-5}$ for the inner, 3.5$\times$ $10^{-4}$ for the intermediate and 3$\times$ $10^{-4}$ for the outer shell. This corresponds to $A_{70\mu m}$ of 7.4$\times$ $10^{-4}$ for all three shells, which for the adopted dust properties/composition (i.e. absorption, and scattering cross sections), translates to a total circumstellar extinction of $A_{V,dust}$ $\sim$ 0.19~($A_{V}$/$A_{70\mu m}$ $\sim$ 257), assuming an averaged grain mass of 2.0$\times$ $10^{-18}$~g. The interstellar extinction can be then determined simply by subtracting the circumstellar extinction from the total extinction derived in Sect.~\ref{dered}, resulting in an A$_{V,IS}$ = 11.8.

\begin{figure*}[ht]
\begin{center}$ 
\begin{array}{cc}
\includegraphics[scale=0.7]{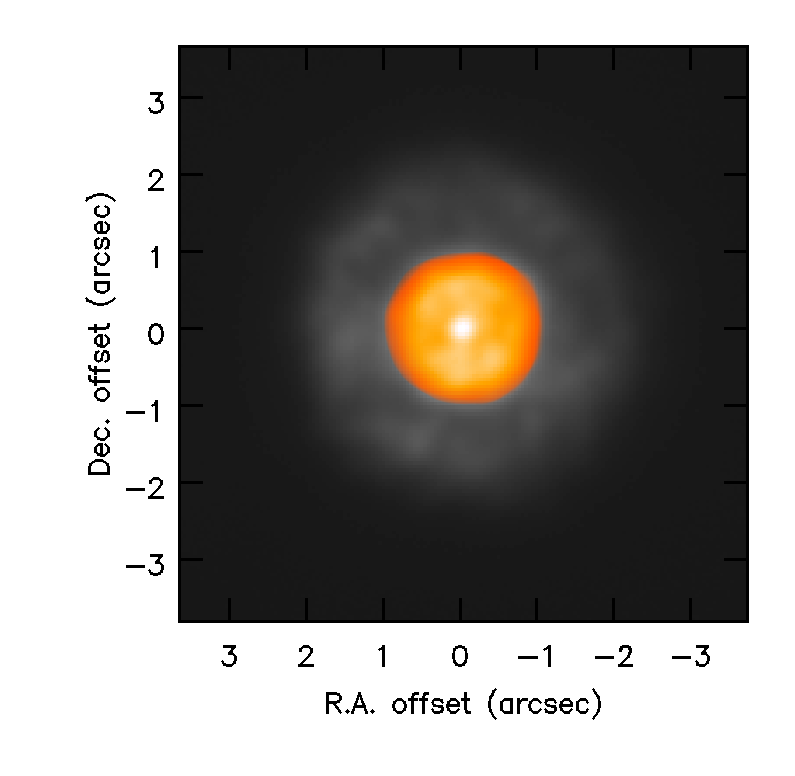} & \includegraphics[scale=0.6]{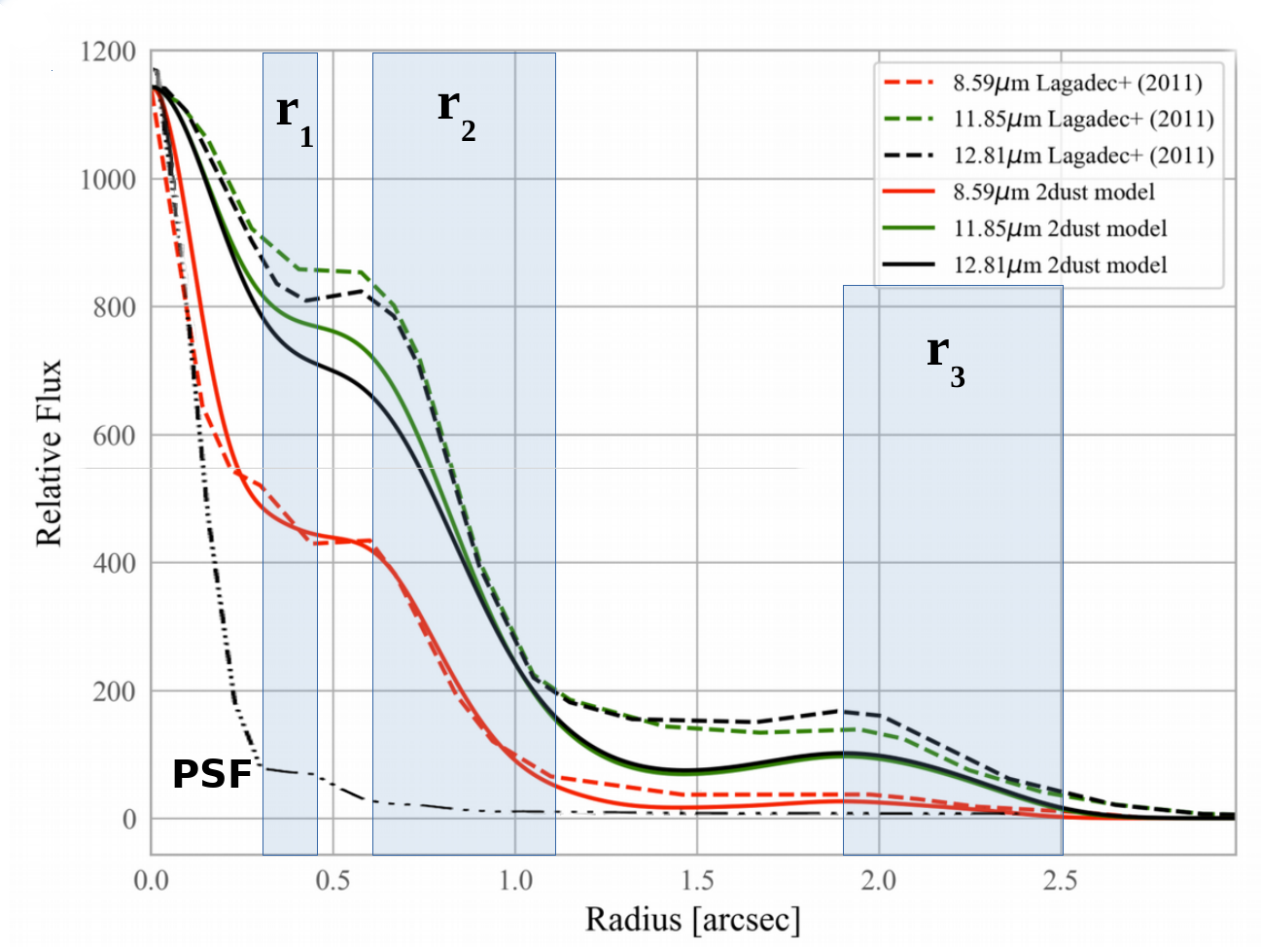} \\
\end{array}$
\end{center}
\caption{{\bf{Left}}: VISIR/VLT mid-infrared monochromatic image at 8.59 $\mu$m \citep{Lagadec2011}. {\bf{Right:}} The radial profiles at 8.59 $\mu$m, 11.85 $\mu$m and 12.81 $\mu$m are extracted and overplotted with the modelled ones as described in Sect.~\ref{2dust}. The size and position of each shell is indicated with light blue shade. The PSF at 11.85 $\mu$m is also overplotted for reference, revealing the hot inner shell. Note that the PSF becomes broader at longer wavelengths.}
\label{visir_images_fit}
\end{figure*}

\begin{figure*}[ht]
\begin{center}  
\includegraphics[scale=0.8]{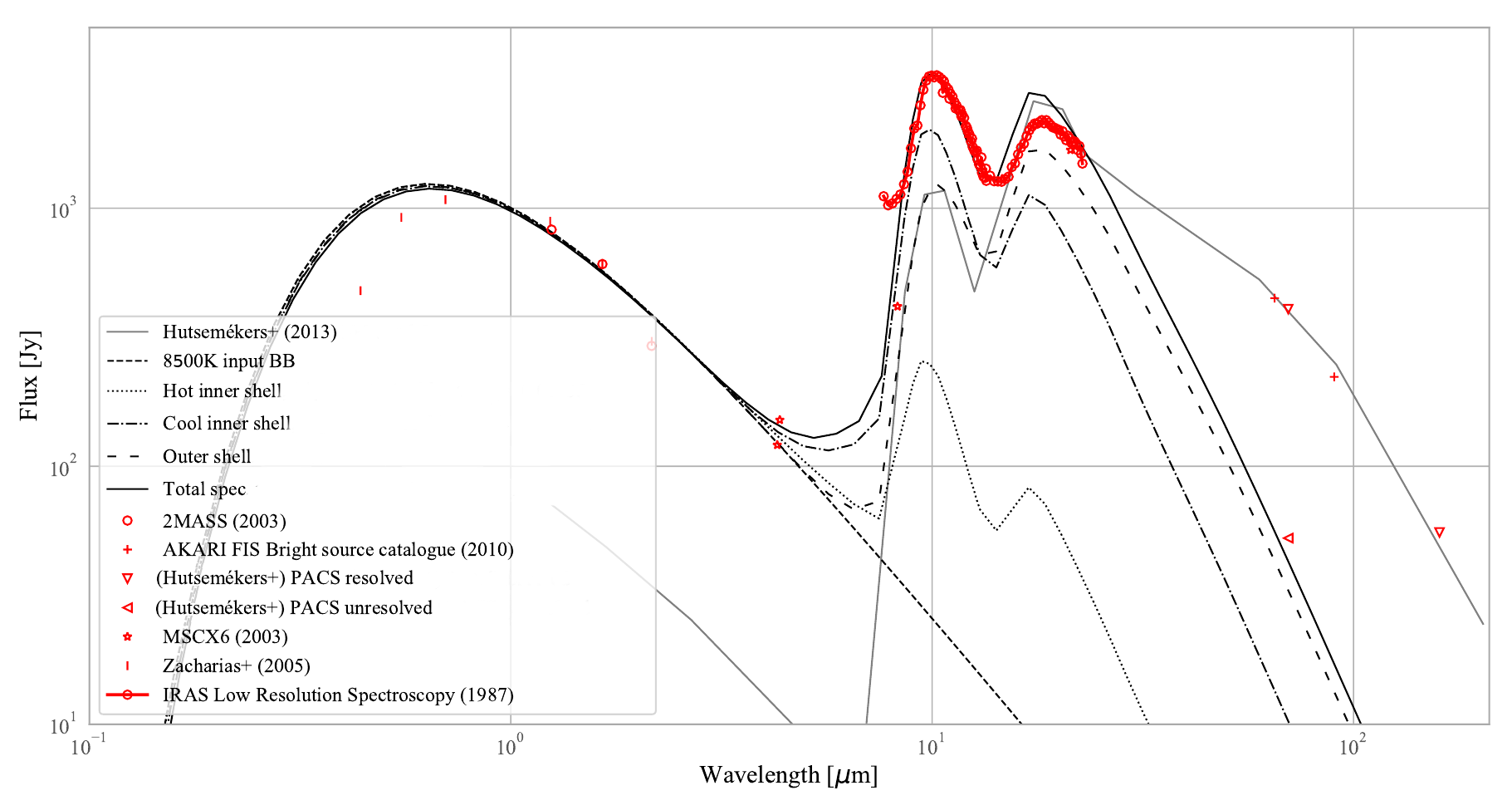} 
\end{center}
\caption{Spectral Energy Distribution fit to the dereddened photometry gathered from literature (symbols). The dashed line shows the 5$\times$10$^{5}$~$\lsun$, 8500 K black-body (BB) input used for the model. The dotted line is the stellar and hot inner shell emission. The dot-dashed line shows the stellar and cool inner shell emission. The large spacing dashed line shows the outer shell emission contribution. We note that our models did not include the ``outermost'' cooler dust shell (18-40") reported in \citet{Hutsemekers2013}, and therefore do not reproduce the far-IR photometry.} 
\label{sed_fit}
\end{figure*}

The SED and intensity radial profile fits are presented in Figures~\ref{visir_images_fit} and~\ref{sed_fit} respectively. In Figure~\ref{sed_fit} we present the resulting SED of the three shells individually (hot inner shell, cool inner shell and outer shell), but also the combined SED (total spec), which was the one we assessed for the goodness of the fit. In the same plot we present the modelled SED presented in \citet{Hutsemekers2013}, which provided a good fit of the IR datapoints (PACS and AKARI). We note that in that work the good fit starts to ``work'' only for the wavelengths beyond 10$\mu$m, showing a very poor fit between 1~$\mu$m and 10~$\mu$m, while the optical photometry below 1~$\mu$m was totally neglected. It is beyond the scope of the current paper to fit the outermost shell of the Fried Egg nebula, and therefore our modelled SED does not reproduce the IR photometry at wavelengths $>$ 70 $\mu m$. Yet given the more complex nature of our models (SED, radial profiles), the dereddening, and the completeness of the photometry, we consider it in overall as an improved fit compared to all previous works. The resulting parameters for each shell are presented in Table~\ref{2dustResults} together with the physical parameters of the central heating source IRAS 17163. 

For the SED dereddening we used the standard interstellar extinction curve \citet{McClure2009}, for a total-to-selective extinction of $R_{V}$ = 3.1, where it is found that extinction effects can be significant at wavelengths as long as 10 $\mu$m, and dereddened the data for the extinction due to the ISM. The extinction curve takes into account the silicate absorption feature from the ISM, for the A$_{V,IS}$ extinction of 11.8, and therefore it is corrected in the SED.

Our synthetic images were produced with the same pixel scale as the 2008 VISIR observations of 0.075\arcsec~per pixel, corresponding to VISIR's pre-upgrade DRS detector. The observed radial profiles of the images presented in \citet{Lagadec2011} can be directly compared to the synthetic radial profiles for a given pixel scale. The number of grid points (radial and latitudinal) was also carefully selected in order to limit the presence of ``artifacts'' in the synthetic maps and extracted radial profiles. To extract the final synthetic radial profile for each  wavelength, we added the resulted images for all three shells, and the central source was included only in the inner shell map. To match the observations the synthetic images were convolved to the angular resolution of the VISIR/VLT ($\theta$ = 1.22~$\lambda$/D), which corresponds to 0.26\arcsec, 0.36\arcsec and 0.39\arcsec at 8.59 $\mu$m, 11.85 $\mu$m and 12.81 $\mu$m respectively for the 8.2~m VLT. The observed radial profiles are asymmetric, with the "left" (eastern) side of the distribution showing some brighter spots compared to the "right" (western) side of the distribution. In this study, for the direct comparison with the perfectly symmetric modelled radial profiles, we "mimic" a perfect symmetry by focusing on the "right side" of the radial profile which is more consistent to the azimuthal averaged intensity profile (Figure~\ref{visir_images_fit}).

The modelled radial profile at 8.59 $\mu$m fits well the observed one at 8.59 $\mu$m, but the fit is less robust with the 11.85 $\mu$m and 12.81 $\mu$m emission. In particular, a larger amount of flux is needed to fit the observed profiles, which becomes more apparent near the location of the inner hot shell. Previous low resolution spectroscopic data \citep[LSR;][]{Epchtein87} were used as a guide for the 10 $\mu$m and 18 $\mu$m silicate features that helped constraining the SED, which is based on observations of 30 years ago. A larger amount of flux at 11.85 $\mu$m and 12.81 $\mu$m, and therefore an improved fit at those wavelengths could possibly be achieved by taking into account a broader silicate feature at 10~$\mu$m, but that is not supported by the LSR data. This led to a compromised fit of those longer wavelengths. Furthermore, the observed images and mass-loss of the Fried Egg nebula are not perfectly symmetric but rather show some clumpiness, which is also expected to contribute to the observed differences. In conclusion, given the above constraints and the sensitivity of the models to the initial dust composition, we consider our three shell model to be able to provide a good fit to the overall shape of both the observed SED and radial profiles.

\subsubsection{Physical properties of the shells}

The sum of the dust mass within the three shells and a canonical gas-to-dust ratio of 100 results in a total mass of $\sim$ $6.5\times 10^{-3}~\msun$ within the Fried Egg Nebula which is about 3 and 1 orders of magnitude lower than what \citet{Lagadec2011} ($\sim$ 4~\msun) and \citet{Hutsemekers2013} ($\sim$0.21~\msun) reported respectively. This difference is mainly due to the consequences of the closer distance we adopt in our study, while the adopted gas-to-dust ratio of 40 in \citet{Hutsemekers2013} could only explain a higher mass by about a factor of 2. The mass of each shell, M$_{d}$, is given by $M_{d} = (4\pi\tau_\lambda r_{d,min} r_{d,max}) / \kappa_\lambda $. Given the closer distance to the object, each shell is also closer to the central heating source by a factor of 3.3 compared to previous works and $\tau_\lambda$ is smaller by a factor of 10. Therefore, the total mass for each shell is expected to be smaller by about 2 orders of magnitude, which confirms what we find.

\begin{table}[ht]
\caption{Physical parameters of the three shells towards the Fried Egg Nebula as modelled using 2-DUST and those of the central heating object IRAS 17163.}
\small
\centering
\setlength\tabcolsep{2pt}
\begin{tabular}{c c c c c c}
\hline\hline

{\bf{Fried Egg Nebula}} & M$_{gas}$ & T$_{d}$ & r$_{d}$ & t$_{kin}$ & $\dot{M}$ \\
& (10$^{-3}$~\msun) & (K) & (\arcsec) & (yr) & (\msun/yr) \\
\hline
Hot inner shell & 0.021 & 620-480 & 0.3-0.45 & 30.8 & 6$\times$10$^{-7}$ \\
Intermediate shell & 0.90 & 460-320 & 0.6-1.1 & 102.7 & 9$\times$10$^{-4}$\\
Outer shell & 5.6 & 240-200 & 1.9-2.5 & 123.2 & 5$\times$10$^{-5}$ \\
\hline\hline
& log(L$\star$/$\lsun$) & T$\star$ & d & & \\
{\bf{IRAS 17163}} & & (K) & (kpc) &  & \\
\hline
 & 5.7 & 8500 & 1.2  &  &\\
\hline\hline

\end{tabular}

\tiny {\bf{Notes}}: The sum of the dust mass within the three shells and a gas-to-dust ratio of 100 results in a total mass of $\sim$6.5~10$^{-3}$~\msun~within the Fried Egg Nebula. We note that the reported total mass does not include the ``outermost'' shell modelled in \citet{Hutsemekers2013}. The reported kinematic scales and mass-loss rates correspond to an expansion velocity, $\upsilon_{exp}$, of 30~km~s$^{-1}$, and therefore provide upper and lower limits respectively.
\label{2dustResults}
\end{table}

Note that both the shells of the Fried Egg Nebula and the larger scale shell are closer to the central heating object than previously thought, due to the closer distance (Sect.~\ref{dist}). The dust temperature is given by $T_{d} = (L\star / 16\pi \sigma r^{2}_d)^{1/4}$, where $L\star$ is the stellar luminosity, $r_d$ is the distance of the dust grain to the star, and $\sigma$ is the Stefan Boltzman constant. Therefore, a physical distance of the shells closer to the star means an increase in $T_{d}$ in the shells compared to what was previously reported.




\section{Discussion}

In this section we discuss some of our main findings, topics of debate in other studies, both theoretical and observational. In particular, we attempt to explain the observed emission spectrum in the K-band, which contains the Na {\sc i} 2.2 $\mu$m doublet in emission, the origin of which has remained unexplained in previous studies. In addition we discuss the varied mass-loss rates of our three modelled distinct shells, which is a very unique finding, that can help to shed more light on both theory and observation of objects in enigmatic stages of stellar evolution.
\label{Discussion}

\subsection{Origin of the Na {\sc i} emission}

\label{NaI_disc}

\subsubsection{Traditional interpretations}

The observed Na {\sc i} 2.2 $\mu$m doublet in emission has been reported in more studies of well-known yellow hypergiants (IRC+10420, HD 179821, HR 8752 and $\rho$~Cas \citep{Lambert1981,Hrivnak1994,Hanson1996,Oudmaijer2013}, LBVs and B[e] stars \citep[e.g.][]{Hamann1986,McGregor1988,Morris1996}. The origin of the Na {\sc i} 2.2 $\mu$m doublet towards evolved massive stellar objects has been traditionally seen as a challenge, and remained unexplained.

We consider several scenarios to explain the origin of this neutral atomic line emission towards IRAS 17163. The neutral Na {\sc i} emission arises from a region that is 10\% larger than the continuum source, and is in turn smaller than the region responsible for the emission component of the Br$\gamma$ profile. \citet{Oudmaijer2013} found a similar situation for IRC+10420. One would expect a neutral metal (Na {\sc i}) to stem from a region further out than that of the ionised hydrogen. Since this is clearly not what we observe, the challenge is to determine which situation can give rise to this apparent contradiction. Here we consider whether Br$\gamma$ emission arises through recombination in a more extended, ionised zone, and Na {\sc i} from the photosphere instead, or whether a mostly neutral atmosphere can account for all the emission lines. 

One of the scenarios is that of a disc. A dusty disc distribution, where both Na {\sc i} and continuum are distributed, allows for shielded Na {\sc i}. The observed more extended Br$\gamma$ emission could then be explained by the penetration of UV/optical radiation at longer distances away from the disc in a low density halo. In Sect.~\ref{size_e} we argued that dust cannot survive at distances $<$ 7~mas, unless it is self-shielded. In that case, one would expect some additional emission from the dust present in the SED at $\sim$ 2.0~$\mu$m, which is not what we see. We conclude that the continuum emission at 2.0~$\mu$m originates from the stellar object directly and thus there are no self-shielded regions close to the star that can provide a place for Na {\sc i} to exist. Therefore, the scenario of a dusty disc distribution at $\sim$ 1~mas (emitting sizes of continuum and Na {\sc i}) can be ruled out. 

Another scenario previously proposed to explain that kind of line emission spectrum is the presence of a pseudophotosphere. \citet{Oudmaijer2013} examine the scenario of an optically thick wind in the case of IRC+10420, which forms a pseudophotosphere and acts as a shielding mechanism for the sodium, preventing its ionisation. This scenario was also supported by \citet{Humphreys2002} based on the high mass-loss rate of $\sim$2$\times$10$^{-4}$ \msun/yr of IRC+10420. As we show below the region within 1.09 - 1.90 mas around the central star cannot be dilute enough for a nebular model to apply and at the same time explain the measured line fluxes. Therefore, this scenario can probably be excluded. 



\subsubsection{A two-zone LTE model}

For a proper interpretation of the 2.1-2.4 $\mu$m emission line spectrum, it turns out not to be necessary to assume that the H Br$\gamma$ line emission arises by recombination of H$^+$ in an ionised gas. In the following we show that a simple model based upon a uniform disk or spherical shell in strict local thermodynamic equilibrium (LTE) reproduces the observations fairly well. Strict LTE means that the ionisation balance is described by the Saha equation for  all the abundant elements and that the energy levels of all these atoms and ions are populated according to the Boltzmann equation at the same temperature. Such a simple model is constructed as follows. Consider a face-on disk (or spherical shell) of radius $r$ and thickness $L=\rho r$ with an angular diameter $\theta = 2r/D$, where $\rho$ is a dimensionless constant and D is the distance to the source. The equilibrium is specified by a set of elemental abundances relative to hydrogen, a temperature $T$, and a number density of hydrogen in all forms (H, H$^+$, H$_2$, H$^-$), $n_{\rm H}$ in cm$^{-3}$. For reference, solar photospheric abundances of the 17 most abundant elements are assumed \citep[cf.][]{Asplund2009}. The abundances of all neutrals, first ions, and a few molecules (H$_2$, CO) are computed in LTE at the specified conditions ($T$ and $n_{\rm H}$). The peak optical depths and flux densities of the infrared lines of interest are then calculated, given the observed linewidths. The continuum opacities and fluxes are computed at the wavelengths of the lines, taking into account electron-ion and electron-neutral free-free opacity (bremsstrahlung), H$^-$ bound-free, and Thomson scattering. A series of such models spanning some range in density and temperature was computed to find the conditions that closely match the observed properties of the Na {\sc i}, Mg {\sc ii}, and H Br$\gamma$ lines in the K band. The adopted observed parameters are summarised in Table~\ref{te}.

\begin{table}[ht]
\caption{K-band observed parameters}
\begin{tabular}{crccc}
\hline
Line & $f_\nu^{obs}$ & $f_\nu^{corr}$ & FWHM & size  \\ 
&  (Jy) & (Jy) &                 km/s  & mas \\
\hline
Na {\sc i} 2.205 &  5.27 &  18.52  &         62 &  1.20 \\
Na {\sc i} 2.208 &  3.17  & 11.11  &        65 &  1.20\\
Br$\gamma$  &  11.20 &  40.85   &       78 &  1.90\\
continuum &  16.21  & 59.14  &        .. &  1.09\\
Mg {\sc ii} 2.136 & 2.34  &  8.76  &        96  &  ..\\
Mg {\sc ii} 2.143 & 1.42  &  5.30  &    & \\
\hline
\end{tabular}
\label{te}
\end{table}

\begin{figure}[ht]
\begin{center}  
\includegraphics[scale=0.45]{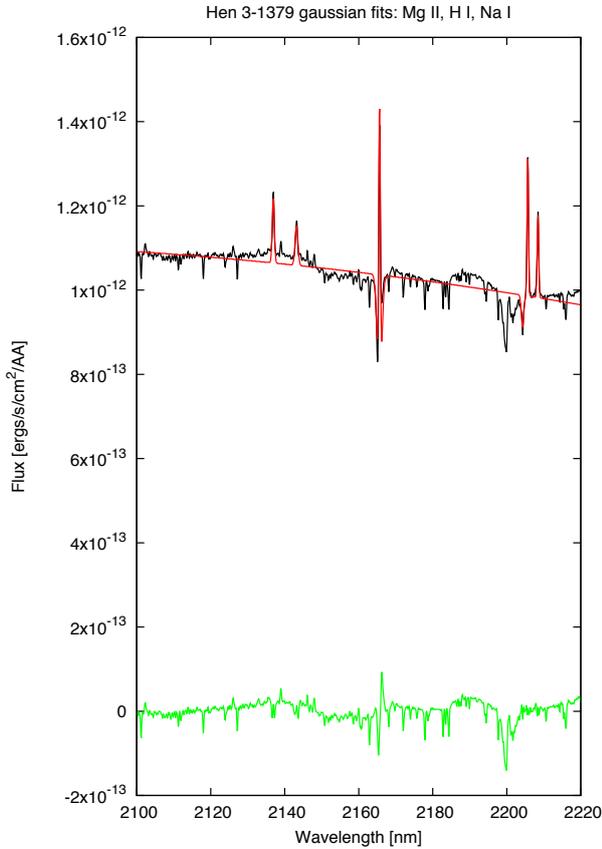} 
\end{center}
\caption{A representative model spectrum in K-band (red) overlaid on the X-Shooter data (black). The model is based on a static spherical shell in strict LTE. The residual spectrum is shown in green.} 
\label{jblack}
\end{figure}

The peak fluxes $f_{\nu}^{\rm obs}$ and line widths (FWHM) were determined from gaussian fits to the X-shooter spectra. A standard extinction curve\footnote{The adopted extinction curve is the renormalised version of the Milky Way dust model with a ratio of total to selective extinction $R_V=3.1$. It was obtained from the website {\tt \url{https://www.astro.princeton.edu/~draine/dust/dustmix.html}}} \citep{Weingartner2001} was used to determine the wavelength-dependent extinction $A_{\lambda}$ appropriate for a visual extinction $A_V=12$ mag. The extinction-corrected flux, $f_{\nu}^{\rm corr} = f_{\nu}^{\rm obs} 10^{0.4 A_{\lambda}}$, is also shown in Table~\ref{te}. The corrected flux is compared directly with the model. To reproduce our observed K-band spectrum we construct a two-zone model as an oversimplification of a dynamical atmosphere. In particular, models with $T\approx 6750$ K, $\log n_{\rm H}=13.2$ cm$^{-3}$, and a ratio of thickness to radius $\rho\approx 0.1$ yield peak flux densities in the near-infrared that are consistent with the observed Na {\sc i} and Mg {\sc ii} emission, for an angular diameter $\theta\approx 1.2$ mas. The peak emission in the Br$\gamma$ profile can be matched with a somewhat more extended region, $\theta\approx 2.0$ mas, at somewhat lower temperature, $T = 5000$ to 5500 K, and density, $\log n_{\rm H} \approx  11.6$ to 12.8 cm$^{-3}$. The optical depth in Br$\gamma$ in this more extended, cooler region is high enough to account for the strong absorption covering the smaller continuum-emitting region, as seen in the P Cygni-type line profile (Fig.~\ref{fig:spectrum3} and \ref{jblack}). Applying expanding wind models with velocity gradients to fully explain the exact shape of the Br$\gamma$ emission line is beyond the scope of the paper. 

Both zones of line emission have low fractional ionisation in LTE, $n(e)/n_{\rm H}\approx 10^{-8}$. Again, we point out that LTE conditions are sufficient to explain the observed emission lines of both neutrals (H {\sc i} and Na {\sc i}) and ions (Mg {\sc ii}) without any need for some kind of photoionised atmosphere (see Fig.~\ref{jblack}). We note that we did not attempt to fit the observed data, and our models point to the range of conditions that can account for the fluxes and angular sizes of the observed infrared emission lines and associated continuum. It also follows that the relative angular sizes of K continuum, Na {\sc i}, and Br$\gamma$ are likely the result of the fact that the optical depth for each varies with radius in the extended atmosphere. The different apparent sizes do not require distinct layers or shells. Br$\gamma$ attains unit optical depth rather farther out in the atmosphere than Na {\sc i}. 

In the future, it would be worthwhile to develop a physically consistent model of the dynamical atmosphere, and to explore additional constraints from the observations (e.g. fit the observed P Cygni-type line profile seen in Br$\gamma$ emission).


The very similar reported mass-loss rates, luminosities and temperatures of IRAS 17163 and IRC+10420, in combination with the presence of a more extended Br$\gamma$ with the coexistence of a more compact Na {\sc i} emission in both cases, make this approach promising in explaining the K-band line emission spectrum towards more objects of this class.


\subsection{Mass-loss}

\label{mass_loss}

\subsubsection{Mass-loss rates}

The consequences of the new distance to the Fried Egg Nebula (smaller optical depth, smaller sizes of the dusty shells) also impacts the derived mass-loss rates. In particular, the model results in a mass-loss rate of 6$\times$10$^{-7}$~\msun~yr$^{-1}$ for the hot inner shell, 9$\times$10$^{-4}$~\msun~yr$^{-1}$ for the intermediate shell and 5$\times$10$^{-5}$~\msun~yr$^{-1}$ for the outer shell.The mass-loss rate in the outer shell is about 2 orders of magnitude lower than what was reported in \citet{Lagadec2011}. 

Large mass-loss rates, higher than 10$^{-5}$~\msun~yr$^{-1}$, can be expected for this class of yellow hypergiants and have been previously measured towards other objects to be $\sim$ 10$^{-4}$~\msun~yr$^{-1}$ \citep[IRC+10420;][]{Oudmaijer1996}, which is comparable (higher by only a factor of 2) to what we find towards the Fried Egg Nebula. In some cases, the mass-loss rates are even as high as $\sim$~5$\times$10$^{-2}$~\msun~yr$^{-1}$ \citep[$\rho$~Cas][]{Lobel2005}. A revision of more yellow hypergiants, where the new Gaia parallaxes are taken into account, is necessary and it may change our current view on mass-loss rates. The mass-loss rate we report for the outer shell is of the same order of magnitude as the one reported in \citet{Wallstrom2017} for the observed parsec scale shell $\sim$~0.8$\times$10$^{-4}$~\msun~yr$^{-1}$. However, these values are the result if one adopts an expansion velocity of 30 km~s$^{-1}$ \citep{Hutsemekers2013} as opposed to 100 km~s$^{-1}$ in \citet{Wallstrom2017}, which differs by a factor of $\sim$~3, influencing the resulting mass-loss rate by the same factor. \citet{Wallstrom2017} estimated the expansion velocity using the CO 2-1 spectrum obtained with ALMA, and in particular by modeling the broad ``plateau'' emission attributing the emission to the stellar wind. Given that we have no information of the situation in the inner shells, being unresolved in \citet{Wallstrom2017}, we also considered the more moderate expansion velocity.

As we already mentioned, the mass-loss rates vary among the three distinct shells. In particular, we find that the intermediate shell has the largest mass-loss rate, followed by the outer shell, while the inner hot shell has the lowest mass-loss rate, which is in fact lower than what it is measured towards other YHGs by about 2 orders of magnitude \citep{Oudmaijer1996}. The fact that the mass-loss rates for the individual shells differ by orders of magnitudes, is something that is probably a real effect that cannot be explained by the uncertainties introduced by the adopted parameters in our models alone (e.g. gas-to-dust ratio, dust composition). \citet{Shenoy2016} reported distinct mass-loss episodes towards a sample of hypergiants, including IRC+10420, at timescales of thousands of years. In particular, IRC+10420 shows also a decline in mass-loss rate up to an order of magnitude, while simulations predict an increase in mass-loss-rate as the object comes closer to exploding as a supernova \citep{Moriya2014}. The physical mechanism responsible for such ``eruptive'' mass-loss events has been traditionally a topic under debate.

\subsubsection{Mass-loss mechanisms}

To explain the presence of shells, we will discuss a pulsational driven mass-loss mechanism and a line driven mass-loss mechanism. In line with previous literature we begin with the classical interpretation of the pulsation hypothesis as the mechanism behind "eruptive" mass-loss events. We then introduce the bi-stability mechanism, which occurs in line-driven winds, and although is well known for hotter objects, it has thus far not been considered in YHG studies. Yet as we discuss, it has the potential to explain the observed mass-loss towards those objects as well.

\vspace{0.5cm}

{\bf{7.2.2.1. Pulsational driven mass-loss}}
\vspace{0.3cm}

YHGs are known to drive significant and unique quasi-period photospheric pulsations \citep[e.g.][]{Lobel1994,deJager1998}, which under favorable conditions \citep[$T_{gas}$ $\sim$ 8000 K;][]{Lobel2001} and on timescales of decades can develop into an eruptive episode. The eruptions are characterised by twice the duration of an ordinary pulsation and by an increased mass-loss rate \citep[e.g. $\rho$~Cas, HR~8752;][]{vanGenderen2019,Lobel2003}. Once hydrogen recombination occurs synchronously with the pulsation decompression (i.e. strong convective motions), results in an outburst event, and a subsequent drop of the temperature \citep{vanGenderen2019}. \citet{LeBertre89} showed a clear variability in the JHK data of IRAS~17163 in a timescale of 3 yrs (up to 0$^{m}$.5 in J) but it was not possible to trace a periodic trend in those data. Given the fact that IRAS~17163 has not been followed up photometrically, it is not possible to identify the observed variability with characteristic patterns of a pulsation mechanism. 

Adopting the stellar parameters (Sect.~\ref{Funda}), the expected pulsation timescale is of order of months \citep[see Eq. 4 in][]{Oudmaijer1994}, while in Sect.~\ref{kin_scales} we find timescales of decades for the shells towards the Fried Egg nebula. We also note that the ejecta caused by the pulsation mechanism in the known objects are optically thick, resulting in a pseudo-photosphere which mimicks a star much cooler than F type. In the case of the Fried Egg nebula, the shells are rather optically thin, and as we discuss in Sect.~\ref{NaI_disc} a pseudo-photosphere is not required to explain the observed emission spectrum. Combining all the information above, we conclude that, if acting, the instabilities leading to the eruptions are less severe in the case of IRAS 17163. We therefore explore another mechanism other than classical pulsations in the following subsection.

\vspace{0.5cm}

{\bf{7.2.2.2. Line driven mass-loss}}
\vspace{0.3cm}

An alternative explanation would involve interacting winds which are driven by radiation pressure in spectral lines. Any change in radiative acceleration has an effect on both the wind velocity and mass-loss rates. If one places IRAS 17163 in the Hertzsprung-Russell diagram, it is located in the white wall/yellow void area. In particular \citet{Nieuwenhuijzen2012} presented two regions, one at $\log$~$T_{eff}$ $\sim$ 3.9-3.95 and one at $\log$~$T_{eff}$ $\sim$ 4.05-4.1. In the first regime where IRAS 17163 is located, the ionisation of H might be responsible for increased dynamical atmospheric instabilities \citep{Humphreys2002}. 

Although one would expect that the most abundant species (hydrogen, helium) are dominating the line force, \citet{Vink1999} showed that more complex ions have a higher contribution because of their very high number of transitions (thousands) in the relevant part of the spectrum where the stellar flux contribution is at its maximum. Iron (Fe) in particular, is a key wind driving element. Little drops in temperatures, result in the recombination of Fe to lower ionisation stages, increasing the density of the wind, reducing its terminal velocity by about a factor of 2 and increasing the mass-loss rate by several factors \citep{Vink1999}. This process is known as bi-stability mechanism. The wind bi-stability is proposed to play a crucial role during shell-ejections \citet{Pauldrach1990}. In particular, these changes are expected at the first bi-stability jump occuring at T $\sim$ 21000~K due to the recombination of Fe {\sc iv} to {\sc iii}, and at the second bi-stability jump occuring at T $\sim$ 8800~K due to the recombination of Fe {\sc iii} to Fe {\sc ii} \citep{Vink2000,Vink2001,Petrov2016}. The evolutionary status and effective temperature of IRAS 17163 - of order 8500 K -, along with the presence of the surrounding multiple, rather optically thin, shells characterised by various mass-loss rates, justify the exploration of the second bi-stability jump. 

The second bi-stability jump predicts escape and terminal wind velocities that are comparable to the outflow velocity of 30-100 km/sec observed towards IRAS 17163 \citep{Lamers1995}. When objects are unstable due to proximity to Eddington limit and approach the opacity bumps, they can enter fast \& slow wind transition cycles, resulting in shells \citep[e.g.,][]{Grafener2012}. If bi-stability is the mass-loss driving mechanism, then the number of the observed shells indicates that IRAS 17163 has undergone four such cycles (including the one by \citep{Hutsemekers2013}).

\subsubsection{Mass-loss geometry}

IRAS 17163 is surrounded by spherical symmetric shells at $\sim$ 0.3$\arcsec$ - 3$\arcsec$ scales \citep[Sec.~\ref{dust_model}; see also,][] {Lagadec2011}. A similar morphology has been seen before in another object of its class, HD~179821 \citep{Ferguson2010}. This morphology has been traditionally explained as mass-loss triggered by a large-scale pulsation mechanism in combination with radiation pressure on the dust. 

As we presented in \ref{channel}, the channel maps (GRAVITY) of the inner regions (mas scales) of IRAS 17163 reveal morphological asymmetries at different velocities. This could be the result of outflowing material in different velocities and orientations, indicating non-uniform mass-loss events at small scales, which may all together lead to a more spherical and symmetric emission. 

At larger scales ($\sim$ 25$\arcsec$), \citet{Wallstrom2017} presented kinematic evidence in CO emission of asymmetric mass-loss and clumpiness connected with the outer shell reported in \citet{Hutsemekers2013}. Asymmetric mass-loss has been previously observed towards other YHGs, with the observed asymmetries attributed to the surface activity of a convective nature \citep[e.g., IRC+10420;][]{Tiffany2010}.

IRAS 17163 was also imaged with the Hubble Space Telescope (HST), but with no sign of a dusty circumstellar shell \citep{Siodmiak2008}. The situation was similar for other YHGs \citep[$\rho$~Cas, HR 8752;][]{Schuster2006}. The lack of nebulosity in HST images has been previously connected to low mass-loss rates \citep[e.g., $\mu$~Cep;][]{deWit2008}, the possible faint nature of the shells at the observed wavelengths or the evolutionary status of the objects \citep{Schuster2006}.


\subsection{Kinematic time-scales}

\label{kin_scales}

The shells are about three times closer to the central star compared to previous studies, and therefore the kinematic time-scales, t$_{kin}$, of each shell have become shorter accordingly. In particular, the time-scales of the intermediate and of the outer shell are found to be 100 and 120 yr respectively. As a consequence, IRAS 17163 has likely passed to its YHG phase more recently than previously thought. The inner hot shell, which we report on and model in this work for the first time, is estimated to be a very recent mass-loss event with a maximum time-scale of only $\sim$~30 yr. It is striking that $\sim$ 30 yrs ago \citet{LeBertre89} presented a clear variability in the JHK data (up to 0$^{m}$.5 in J) towards the object. Given our new findings, this variability can very well be the result of an eruption that led to the formation of the inner third shell. It is worth mentioning that at the updated distance of 1.2 kpc the timescale of the cold, more extended, outer shell presented in \citet{Hutsemekers2013} is roughly reduced from 16000 yr to 4000 yr. Our reported values are most likely an upper limit as they were derived for the lower expansion velocity of 30 km/s. For example, if we would use 100 km/s as done by \citet{Wallstrom2017}, instead, the values would be shorter by a factor of 3. With kinematic time-scales of 100 yr or shorter, the star is expected to have evolved notably over the last century, something that could be traced with changes also in the visual magnitude or the infrared fluxes. For future reference, it would be worth investigating if there are any archival\footnote{We investigated the Harvard College Observatory plate archive (available at http://dasch.rc.fas.harvard.edu/project.php), traces back to approximately 100 years of photometric observations, and the archive of the American Association of Variable Star Observers (https://www.aavso.org/) and found no long-term photometric measurements of IRAS 17163.}, historical photometric measurements to trace the long-term variability of IRAS 17163.

We conclude that the Fried Egg Nebula is a key object which is characterised by three distinct mass-loss events with varied mass-loss rates and maximum timescales from 30 yr up to 120 yr, with the most recent event to be the less powerful in terms of mass-loss compared to the previous two. Given the observational restrictions, IRAS 17163 is among the very few cases where multiple mass ejections have been reported at such sort timescales \citep[see also $\rho$~Cas;][]{Beardsley1961,Shenoy2016}. Such objects can shed light on fundamental questions on massive star evolution such as how many repeating eruptions can be expected to occur for a particular object, what is the exact cycle of the eruptions, and what are the dependencies between the mass-loss rates and episodes and the local conditions (e.g. mass, effective temperature, clumpiness, metallicity).

\section{Summary}

\label{conclusions}

This paper presents new optical and near infrared spectroscopy (X-Shooter) and high spatial resolution (down to mas) GRAVITY and AMBER observations of the yellow hypergiant IRAS 17163. In particular we reported on the spectral type of IRAS 17163, the updated distance using the new Gaia measurements and the geometry of the continuum emission and that of Br$\gamma$ and Na {\sc i} line emission close to the object, image reconstruction, but also on modeling published photometry by applying the radiative transfer code 2-Dust. 

Evolved stars, with their high luminosity and extended atmospheres are great targets for optical/near infrared interferometry, but yet their imaging is challenging, with only few objects known to date to be meeting the criteria for a successful image reconstruction at these wavelengths \citep[e.g. $\eta$ Car,][Gravity Collaboration]{Weigelt2016,Sanchez2018}. We present the first image reconstruction of IRAS 17163 at short wavelengths ($\sim$2~$\mu$m) and provide new insights into its close surroundings but also on the characteristics of its surrounding shells and mass-loss episodes.

To interpret the observed spherical shells we explored both pulsational driven mass-loss and line driven mass-loss mechanisms. We conclude that with only one object and the observations in hand we cannot clearly discriminate among the two mechanisms. The bi-stability mechanism with its second bi-stability jump provides a promising and unified mechanism to explain mass-loss processes towards both hot (blue S Dor variables; first bi-stability jump) and cooler (YHGs; second bi-stability jump) stages of stellar evolution. A future study where the pros and cons of both the pulsation and bi-stability mechanisms are thoroughly discussed towards a sample of all known YHGs will hopefully allow to properly distinguish among the two theories on the variable mass-loss at those stages. 

The key results of our study, which is based on optical and near infrared spectroscopy and high spatial resolution observations of the yellow hypergiant IRAS 17163, are presented as follows.

\begin{itemize}
\item Based on classic spectral typing (between A3 and A6) and the analogy with IRC +10420 we infer the photospheric temperature of IRAS 17163 to be $\sim$ 8500 K. 
\item The continuum emission at 2~$\mu$m stems from the star and is not associated with hot dust.
\item Our models reveal the existence of a third inner hot shell for the first time. This is a very recent mass-loss event with a time-scale of maximum 30 yr.
\item The three distinct shells trace three mass-loss episodes and are all characterised by variability in the mass-loss rate, with the lowest mass-loss rate to be that of the inner shell. 
\item The GRAVITY model-independent image reconstruction at the selected wavelengths (Br$\gamma$, Na {\sc i} and continuum) shows the more extended and rather asymmetric nature of the Br$\gamma$ emission (E-W direction; e.g. asymmetric wind origin) compared to Na {\sc i} and continuum. In particular, the Br$\gamma$ shows also a northern and southern component, while Na {\sc i} appears to follow the distribution of the photosphere. A similar situation has previously been observed by applying geometrical models towards another object of the same class (IRC+10420). Our spectropolarimetric analysis betrays asymmetries of order stellar radii along an EW position angle, which is in agreement with the proposed structure.
\item The Br$\gamma$ emission shows a P-Cygni profile and a drop in its peak intensity up to $\sim$~40\% in a four month period, indicative of a variable wind from the central YHG with variable gas mass-loss rates.
\item The reported mass-loss rates and kinematic time-scales of the shells constituting the Fried Egg Nebula, are consistent with the predictions from the bi-stability mechanism and not those predicted by the classical pulsation hypothesis.
\end{itemize}

\begin{acknowledgements}
EK is funded by the STFC (ST/P00041X/1). We thank Sundar Srinivasan, Pierre Kervella and Werner Salomons for their help when preparing the paper and useful discussions. We also thank both referees for a careful reading of the manuscript and for providing useful comments and suggestions that improved the paper. This research has made use of the \texttt{AMBER data reduction package} of the Jean-Marie Mariotti Center\footnote{Available at http://www.jmmc.fr/amberdrs}.
Based on observations collected at the European Southern Observatory
under ESO programme(s) 099.D-141 (X-Shooter), 089.D-0576 (AMBER), 099.D-0664 (GRAVITY). This research has made use of the SIMBAD data base, operated at CDS, Strasbourg, France. This research has made use of the Jean-Marie Mariotti Center \texttt{OImaging}
service \footnote{Available at http://www.jmmc.fr/oimaging} part of the European Commission's FP7 Capacities programme (Grant Agreement Number 312430). The GRAVITY data reduction presented in this work was undertaken using ARC3, part of the High Performance Computing facilities at the University of Leeds, UK.
\end{acknowledgements}

\bibliographystyle{aa} 
\bibliography{ref}


\begin{appendix}

\section{Observations}

\begin{figure*}[ht]
  \begin{center}
\includegraphics[scale=0.6]{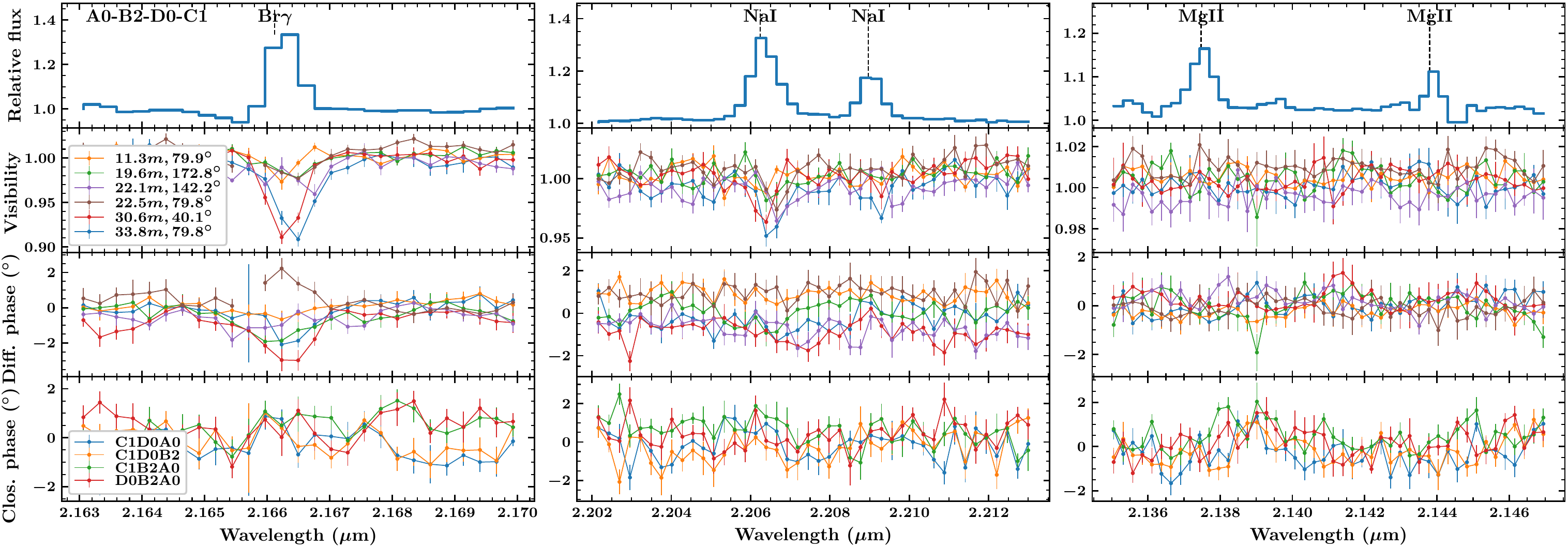}
  \end{center}
\caption{As Fig.~\ref{fig:vis_brg5}, but for the small configuration A0-B2-C1-D0. The drop in visibilities at $\sim$~2.139~$\mu$m is based on a single data point and not connected to an emission line, therefore we consider it an artifact. The closure phase does not show changes in this configuration.}
\label{fig:vis_brg1_2}
\end{figure*}

\begin{figure*}[ht]
\begin{center}  
\includegraphics[scale=0.6]{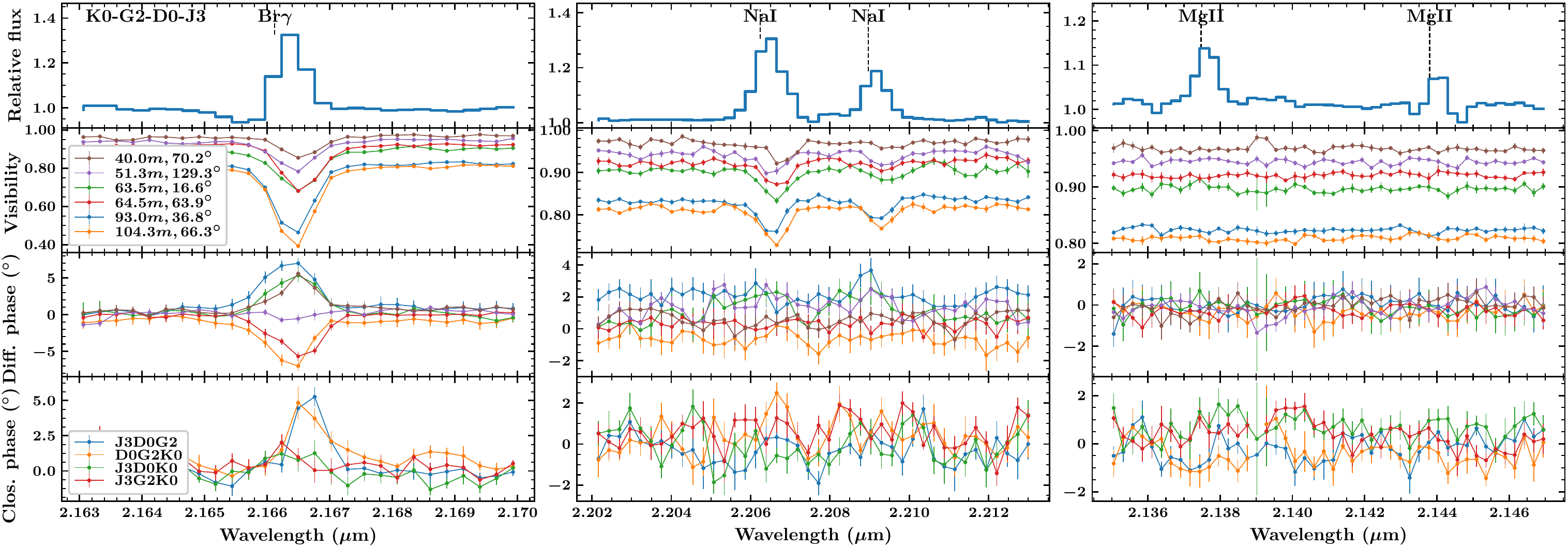} 
\end{center}
\caption{As Fig.~\ref{fig:vis_brg5}, but for the medium configuration K0-G2-D0-J3. Note that the changes in differential phases from positive to negative values, correspond to similar baselines but different PA by $>$30 degrees, and are indicative of offsets of the photocentre projected in a specific baseline.}
\label{fig:vis_brg2}
\end{figure*}

\begin{figure*}[ht]
\begin{center}  
\includegraphics[scale=0.6]{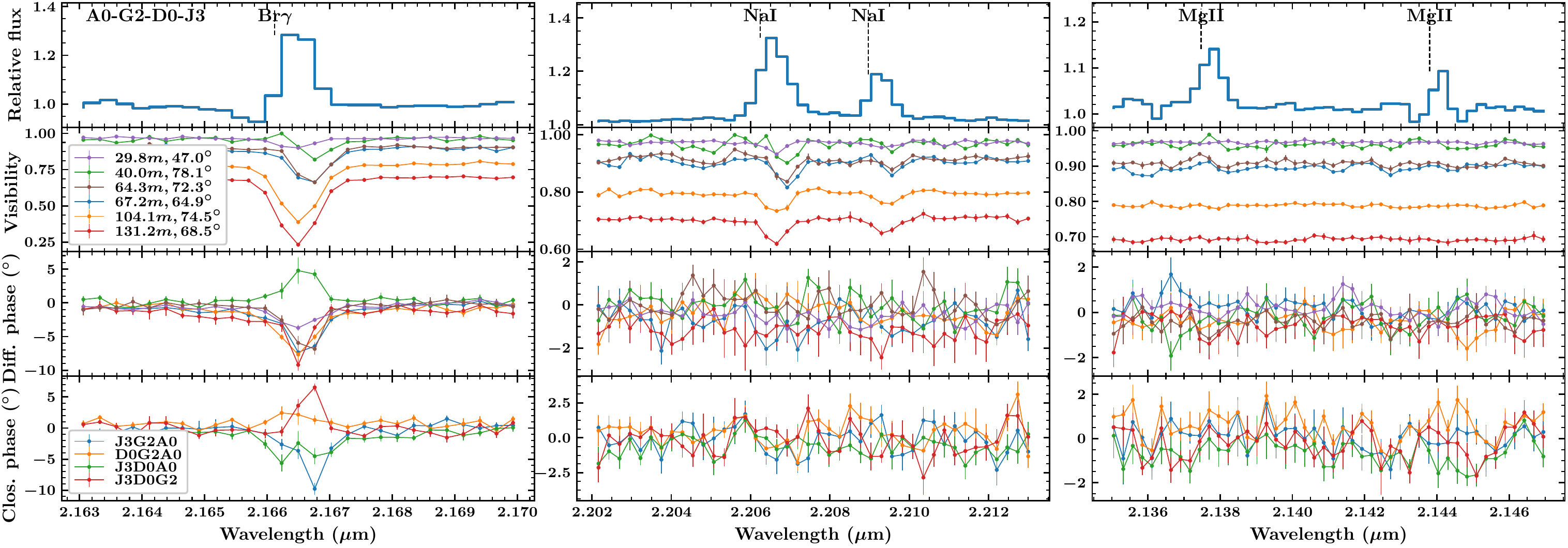} 
\end{center}
\caption{As Fig.~\ref{fig:vis_brg5}, but for the medium configuration A0-G2-D0-J3.}
\label{fig:vis_brg3}
\end{figure*}

\begin{figure*}[ht]
\begin{center}  
\includegraphics[scale=0.6]{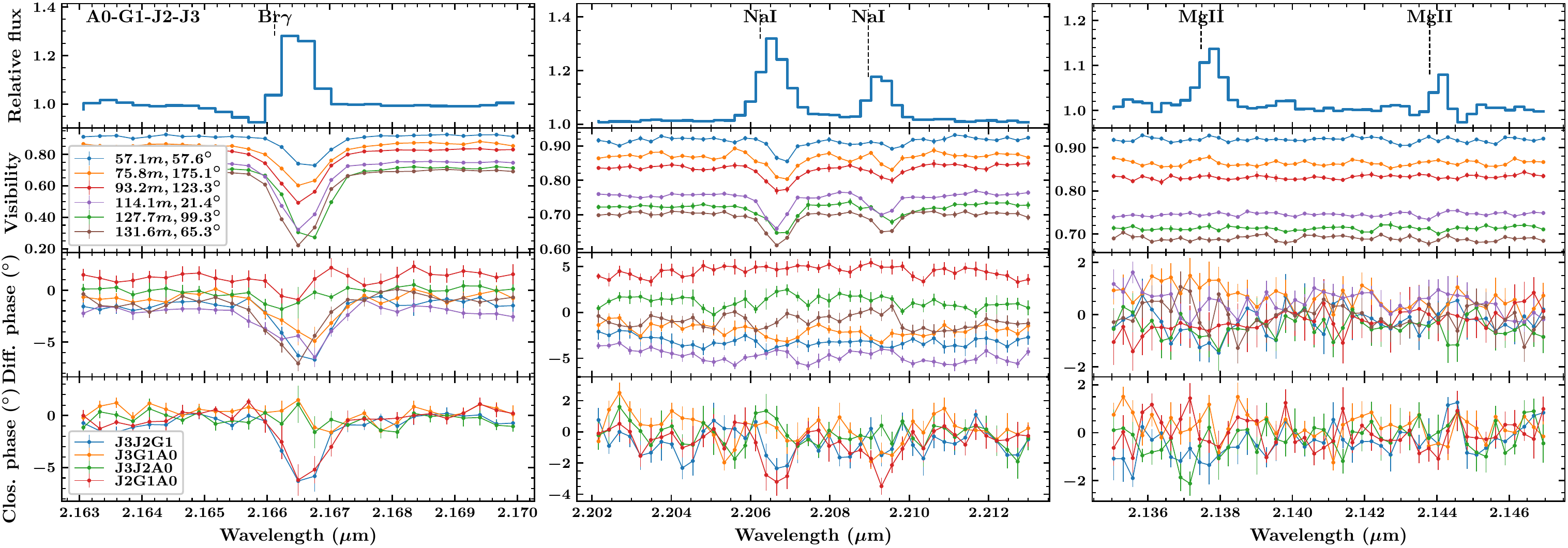} 
\end{center}
\caption{As Fig.~\ref{fig:vis_brg5}, but for the large configuration A0-G1-J2-J3. This is the only case where Na {\sc i} also shows a change in the closure phase.}
\label{fig:vis_brg4}
\end{figure*}

\clearpage

\begin{table*}[ht]
\caption{Technical overview of the GRAVITY observations of IRAS 17163}
\small
\centering
\setlength\tabcolsep{2pt}
\begin{tabular}{c c c c c c c c c c c c c c c}
\hline\hline
Config. & Date & Station & Baseline & PA & N$\times$DIT & $\tau_{coh}$ & Seeing & V$_{cont}$ & V$_{Br\gamma}$ & V$_{Na I}$  \\  &  & & (m) & ($^\circ$) & (s) & (ms) & (arcsec) & & & \\
\hline\hline
A0-B2-D0-C1 & 2017-04-11 & C1B2 & 11.3 & 79.8 & 10$\times$30 & 5 & 0.80 & 1.004$\pm$0.001 & 1.012$\pm$0.006 & 0.988$\pm$0.009 \\
&  & C1A0 & 19.6 & 172.8 &  &  &  & 1.0035$\pm$0.0007 & 0.993$\pm$0.005 & 0.981$\pm$0.006 \\
&  & B2A0 & 22.1 & 142.2 &  &  &  & 0.9981$\pm$0.0007 & 0.959$\pm$0.011 & 0.989$\pm$0.005 \\
&  & C1D0 & 22.5 & 79.8 &  &  &  & 1.0104$\pm$0.0005 & 0.989$\pm$0.004 & 0.990$\pm$0.009\\
&  & D0A0 & 30.6 & 40.1 &  &  &  & 1.0035$\pm$0.0005 & 0.991$\pm$0.005 & 0.963$\pm$0.004 \\
&  & D0B2 & 33.8 & 79.8 &  &  &  & 0.998$\pm$0.002 & 0.951$\pm$0.008 & 0.951$\pm$0.006\\
A0-B2-D0-C1 & 2017-04-12 & C1B2 & 11.0 & 62.7 & 10$\times$30 & 3 & 0.94 & 0.83$\pm$0.01 & 0.8$\pm$0.2 & 0.7$\pm$0.2\\
&  & C1D0 & 22.0 & 62.7 &  &  &  & 0.4633$\pm$0.0007 & 0.425$\pm$0.008 & 0.405$\pm$0.006 \\
&  & C1A0 & 22.5 & 154.0 &  &  &  & 0.97$\pm$0.02 & 0.9$\pm$0.2 & 0.9$\pm$0.2\\
&  & B2A0 & 24.8 & 127.8 &  &  &  & 0.86$\pm$0.02 & 0.8$\pm$0.2 & 0.8$\pm$0.2 \\
&  & D0A0 & 31.9 & 17.7 &  &  &  & 0.464$\pm$0.007 & 0.42$\pm$0.09 & 0.38$\pm$0.1 \\
&  & D0B2 & 33.0 & 62.7 &  &  &  & 0.428$\pm$0.007 & 0.37$\pm$0.09 & 0.3$\pm$0.1 \\
A0-B2-D0-C1 & 2017-06-18 & C1B2 & 10.5 & 53.9 & 10$\times$30 & 2 & 0.66 & 0.961$\pm$0.001& 0.96$\pm$0.01 & 0.974$\pm$0.007\\
&  & C1D0 & 21.0 & 53.8 &  &  &  & 0.970$\pm$0.001 & 0.95$\pm$0.01 &0.976$\pm$0.008 \\
&  & C1A0 & 22.4 & 140.6 &  &  &  & 0.951$\pm$0.001 & 0.93$\pm$0.01 & 0.95$\pm$0.01\\
&  & B2A0 & 25.3 & 116.2 &  &  &  & 0.9708$\pm$0.0005 & 0.947$\pm$0.008 & 0.964$\pm$0.003\\
&  & D0A0 & 29.9 & 5.3 &  &  &  & 0.969$\pm$0.001 & 0.92$\pm$0.01 & 0.95$\pm$0.02\\
&  & D0B2 & 31.5 & 53.8 &  &  &  & 0.960$\pm$0.001 & 0.89$\pm$0.01 & 0.95$\pm$0.01\\
\hline
K0-G2-D0-J3 & 2017-04-22 & D0G2 & 40.0 & 71.9 & 10$\times$30 & 9 & 0.53 & 0.9652$\pm$0.0004 & 0.878$\pm$0.003 & 0.963$\pm$0.003\\
&  & J3K0 & 50.7 & 130.2 & & & & 0.9547$\pm$0.0005 & 0.870$\pm$0.004 & 0.915$\pm$0.007\\
&  & G2K0 & 62.6 & 18.7 &  &  &  & 0.9033$\pm$0.0005 & 0.746$\pm$0.005 & 0.872$\pm$0.005\\
&  & J3G2 & 64.5 & 65.7 & & & & 0.9215$\pm$0.0003 & 0.729$\pm$0.005 & 0.888$\pm$0.004\\
&  & D0K0 & 92.2 & 39.0 &  &  &  & 0.8314$\pm$0.0005 & 0.639$\pm$0.004 & 0.75$\pm$0.01 \\
&  & J3D0 & 104.3 & 68.1 &  &  &  & 0.8116$\pm$0.0005 & 0.571$\pm$0.006 & 0.750$\pm$0.003\\
K0-G2-D0-J3 & 2017-04-24 & D0G2 & 39.3 & 56.5 & 10$\times$30 & 5 & 0.39 & 0.60$\pm$0.01 & 0.55$\pm$0.07 & 0.6$\pm$0.1\\
&  & J3K0 & 54.8 & 119.2 &  &  &  & 0.636$\pm$0.007 & 0.55$\pm$0.07 & 0.58$\pm$0.08\\
&  & J3G2 & 63.7 & 49.4 &  &  &  & 0.30$\pm$0.01 & 0.2$\pm$0.1 & 0.3$\pm$0.1 \\
&  & G2K0 & 68.3 & 0.4 &  &  &  & 0.16$\pm$0.01 & 0.14$\pm$0.09 & 0.16$\pm$0.10\\
&  & D0K0 & 95.9 & 20.3 &  &  &  & 0.13$\pm$0.01 & 0.1$\pm$0.1 & 0.1$\pm$0.1\\
&  & J3D0 & 102.8 & 52.1 &  &  &  & 0.11$\pm$0.01 & 0.07$\pm$0.10 & 0.1$\pm$0.1\\
\hline
A0-G2-D0-J3 & 2017-04-25 & D0A0 & 29.8 & 47.0 & 10$\times$30 & 5 & 0.60 & 0.9678$\pm$0.0004 & 0.931$\pm$0.004 & 0.9542$\pm$0.004\\
&  & D0G2 & 40.0 & 78.1 &  &  &  & 0.9621$\pm$0.0005 & 0.819$\pm$0.003 & 0.995$\pm$0.008\\
&  & J3G2 & 64.3 & 72.3 &  &  &  & 0.9074$\pm$0.0007 & 0.662$\pm$0.009 & 0.919$\pm$0.007 \\
&  & G2A0 & 67.2 & 64.9 &  &  &  & 0.8949$\pm$0.0004 & 0.663$\pm$0.003 & 0.902$\pm$0.008\\
&  & J3D0 & 104.1 & 74.5 &  &  &  & 0.7884$\pm$0.0005 & 0.497$\pm$0.004 & 0.745$\pm$0.007\\
&  & J3A0 & 131.2 & 68.5 &  &  &  & 0.6943$\pm$0.0007 & 0.381$\pm$0.009 & 0.64$\pm$0.01\\
\hline
A0-G1-J2-J3 & 2017-04-27 & J2G1 & 56.9 & 30.5 & 10$\times$30 & 6 & 0.45 & 0.9051$\pm$0.0003 & 0.741$\pm$0.005 & 0.891$\pm$0.003 \\
&  & G1A0 & 90.4 & 150.9 &  &  &  & 0.8605$\pm$0.0002 & 0.590$\pm$0.006 & 0.831$\pm$0.004\\
&  & J3J2 & 101.6 & 106.6 &  &  &  & 0.8193$\pm$0.0004 & 0.539$\pm$0.002 & 0.778$\pm$0.003\\
&  & J3G1 & 127.8 & 81.0 &  &  &  & 0.6963$\pm$0.0004 & 0.278$\pm$0.005 & 0.674$\pm$0.004 \\
&  & J3A0 & 128.7 & 39.7 &  &  &  & 0.7072$\pm$0.0003 & 0.388$\pm$0.006 & 0.639$\pm$0.006\\
&  & J2A0 & 128.9 & 173.3 &  &  &  & 0.7134$\pm$0.0003 & 0.373$\pm$0.003 & 0.643$\pm$0.004\\
A0-G1-J2-J3 & 2017-04-28 & J2G1 & 56.9 & 59.7 & 10$\times$30 & 4 & 0.48 & 0.9145$\pm$0.0004 & 0.734$\pm$0.005 & 0.909$\pm$0.006\\
&  & G1A0 & 73.9 & 176.8 &  &  &  & 0.8678$\pm$0.0003 & 0.648$\pm$0.007 & 0.852$\pm$0.003\\
&  & J3J2 & 92.2 & 124.2 &  &  &  & 0.8305$\pm$0.0004 & 0.569$\pm$0.003 & 0.798$\pm$0.003\\
&  & J2A0 & 111.9 & 23.7 &  &  &  & 0.7541$\pm$0.0003 & 0.426$\pm$0.003 & 0.702$\pm$0.004\\
&  & J3G1 & 127.5 & 100.5 &  &  &  & 0.7125$\pm$0.0005 & 0.275$\pm$0.004 & 0.689$\pm$0.008\\
&  & J3A0 & 131.4 & 67.3 &  &  &  & 0.6905$\pm$0.0005 & 0.333$\pm$0.009 & 0.645$\pm$0.004\\
A0-G1-J2-J3 & 2017-08-28 & J2G1 & 51.9 & 19.7 & 10$\times$30 & 4 & 0.56 & 0.9302$\pm$0.0002 & 0.816$\pm$0.006 & 0.935$\pm$0.001\\
&  & G1A0 & 88.8 & 136.8 &  &  &  & 0.8389$\pm$0.0004 & 0.623$\pm$0.007 & 0.822$\pm$0.004\\
&  & J3J2 & 102.8 & 94.8 &  &  &  & 0.8175$\pm$0.0005 & 0.542$\pm$0.003 & 0.815$\pm$0.005 \\
&  & J3A0 & 118.1 & 29.8 &  &  &  & 0.7059$\pm$0.0003 & 0.389$\pm$0.002 & 0.700$\pm$0.007\\
&  & J2A0 & 121.6 & 159.1 &  &  &  & 0.7111$\pm$0.0002 & 0.387$\pm$0.002 & 0.695$\pm$0.004\\
&  & J3G1 & 125.3 & 72.5 &  &  &  & 0.7277$\pm$0.0004 & 0.384$\pm$0.005 & 0.735$\pm$0.005 \\
\hline
A0-G1-J2-K0 & 2017-07-26 & K0J2 & 47.3 & 82.5 & 10$\times$30 & 4 & 0.76 & 0.9598$\pm$0.0005 & 0.840$\pm$0.005 & 0.957$\pm$0.005\\
&  & J2G1 & 47.5 & 14.1 &  &  &  & 0.9559$\pm$0.0007 & 0.820$\pm$0.005 & 0.96$\pm$0.01 \\
&  & K0G1 & 78.4 & 48.2 &  &  &  & 0.8976$\pm$0.0005 & 0.63$\pm$0.01 & 0.886$\pm$0.007\\
&  & G1A0 & 86.6 & 128.1 &  &  &  & 0.8679$\pm$0.0009 &0.597$\pm$0.006 & 0.848$\pm$0.008\\
&  & K0A0 & 106.2 & 174.8 &  &  &  & 0.7920$\pm$0.0005 & 0.437$\pm$0.005 & 0.775$\pm$0.007\\
&  & J2A0 & 114.5 & 150.4 &  &  &  & 0.7669$\pm$0.0005 & 0.385$\pm$0.006 & 0.748$\pm$0.005\\
\hline\hline
\end{tabular}
\label{tech}
\end{table*}

\begin{table*}[ht]
\caption{Technical overview of the X-Shooter observations of IRAS 17163}
\begin{tabular}{crc}
\hline
Date Obs. & t$_{\rm exp}$ (s) & $\lambda$ range (nm )   \\ 
\hline
2017-04-08 & 2$\times$160   & 300 - 560                    \\ 
2017-04-14 & 2$\times$160   & 300 - 560                  \\ 
2017-05-29 & 2$\times$1060  & 300 - 560                  \\ 
2017-05-30 & 2$\times$1286, 2$\times$940        & 300 - 560                  \\ 
2017-04-08 & 3$\times$5   & 560 - 1020                   \\ 
2017-04-14 & 3$\times$5   & 560 - 1020                  \\ 
2017-05-29 & various, 19$\times$45                  \\ 
2017-05-30 & various, 13$\times$60       & 560 - 1020                   \\ 

\hline
\end{tabular}

\tiny {\bf{Notes}}:Usable wavelength range, SNR larger than 5. SNR measured in line free regions around 660 nm. The VIS X-Shooter data were taken with various exposures, ranging from 1, 2, 5, 10, to 45s in order to avoid saturation in bright sections. Only the longest exposures are listed. NIR data were taken with the shortest exposures, but few were usable. 
\label{logspec}
\end{table*}

\begin{table*}[ht]
\caption{Technical overview of the AMBER observations of IRAS 17163 during the observing night of 2012-06-22, where DIT is the individual exposure time, and $\tau_{coh}$ is the coherence time.}
\small
\centering
\setlength\tabcolsep{2pt}
\begin{tabular}{c c c c c c c c c c c c c}
\hline\hline
Config. & Station & Baseline & PA & DIT & $\tau_{coh}$ & Seeing & V$_{cont}$ & V$_{Br\gamma}$  \\ & & (m) & ($^\circ$) & (s) & (ms) & (arcsec) & & \\
\hline\hline
D0-I1-H0 & D0I1 & 64.0 & 23.3 & 6 & 3 & 1.2 & 0.86$\pm$0.62 & 0.36$\pm$0.28 \\
& I1H0 & 32.6 & 139.1 &  &  &  & 0.89$\pm$0.49 & 0.68$\pm$0.37 \\
& D0H0 & 57.8 & 53.8 &  &  &  & 0.92$\pm$0.43 & 0.62$\pm$0.30 \\
& D0I1 & 70.1 & 15.1 & 6  & 3 & 1.2 & 0.86$\pm$0.46 & 0.64$\pm$0.20 \\
& I1H0 & 34.9 & 136.2 &  &  &  & 0.91$\pm$0.34 & 0.73$\pm$0.62 \\
& D0H0 & 60.0 & 44.9 &  &  &  & 0.98$\pm$0.37 & 0.75$\pm$0.32 \\
& D0I1 & 76.7 & 5.7 &6  & 3 & 1.3 & 0.88$\pm$0.44 & 0.54$\pm$0.17 \\
& I1H0 & 37.3 & 131.6 &  &  &  & 1.02$\pm$0.43 & 0.80$\pm$0.33 \\
& D0H0 & 62.5 & 34.6 &  &  &  & 0.87$\pm$0.52 & 0.5$\pm$0.35 \\
D0-H0-G1 & D0H0 & 63.0 & 14.0 & 6  & 3 & 1.1 & 0.84$\pm$0.49 & 0.64$\pm$0.50 \\
 & H0G1 & 69.0 & 79.0 &  &  &  & 0.91$\pm$0.61 & 0.55$\pm$0.44 \\
 & D0G1 & 71.1 & 132.4 &  &  &  & 0.94$\pm$0.63 & 0.37$\pm$0.30  \\
 & D0H0 & 52.1 & 173.5 & 6  & 4 & 1.0 & 0.88$\pm$0.36 & 0.61$\pm$0.22  \\
 & H0G1 & 65.9 & 66.5 &  &  &  & 0.86$\pm$0.40 & 0.45$\pm$0.17  \\
 & D0G1 & 71.0 & 111.0 &  &  &  & 0.90$\pm$0.57 & 0.71$\pm$0.24  \\
 & D0H0 & 44.6 & 163.1 & 6  & 3 & 1.4 & 0.89$\pm$0.49 & 0.81$\pm$0.37  \\
 & H0G1 & 63.4 & 62.4 &  &  &  & 0.81$\pm$0.54 & 0.55$\pm$0.22  \\
 & D0G1 & 70.4 & 100.9 &  &  &  & 0.75$\pm$0.54 & 0.63$\pm$0.38  \\
 & D0H0 & 36.5 & 149.7 & 6  & 5 & 0.7 & 0.93$\pm$0.55 & 0.88$\pm$0.48  \\
 & H0G1 & 60.3 & 59.2 &  &  &  & 0.82$\pm$0.30 & 0.61$\pm$0.41  \\
 & D0G1 & 70.2 & 90.5 &  &  &  & 0.85$\pm$0.46 & 0.40$\pm$0.23  \\
\hline\hline
\end{tabular}
\label{amber}
\end{table*}

\begin{table*}[ht]
\caption{Photometry of IRAS 17163}
\begin{tabular}{clccc}
\hline
Filter & Magnitude  & Flux$_{obs}$ & Flux$_{dered}$ & Reference  \\
 & (mag) & (Jy) & (Jy) &   \\
\hline
B &  17.13  & 6$\times$10$^{-4}$ & 1.54$\times$10$^{3}$ & \citet{LeBertre89}   \\  
V &  13.03  & 2.35$\times$10$^{-2}$ & 1.47$\times$10$^{3}$ & \citet{LeBertre89}    \\ 
J &  4.635  & 2.23$\times$10$^{1}$ & 5.01$\times$10$^{2}$ & \citet{Cutri2003}  \\
H &  3.021  & 6.34$\times$10$^{1}$ & 5.18$\times$10$^{2}$ & \citet{Cutri2003}  \\
K &  2.407  & 7.26$\times$10$^{1}$ & 2.56$\times$10$^{2}$ & \citet{Cutri2003} \\ 
L &  1.820  & 5.86$\times$10$^{1}$ & 1.09$\times$10$^{2}$ & \citet{Epchtein87}   \\              

\hline
\end{tabular}

\tiny {\bf{Notes}}: The observed flux was dereddened using $A_{V}$ = 12.00 as described in Sect.~\ref{dered}.

\label{photometry_data}
\end{table*}

\clearpage

\section{Image reconstruction}
\subsection{Channel maps}
\label{channel}

\begin{figure*}[h]
\begin{center}
\includegraphics[scale=0.85,trim={0 550 0 0},clip=true]{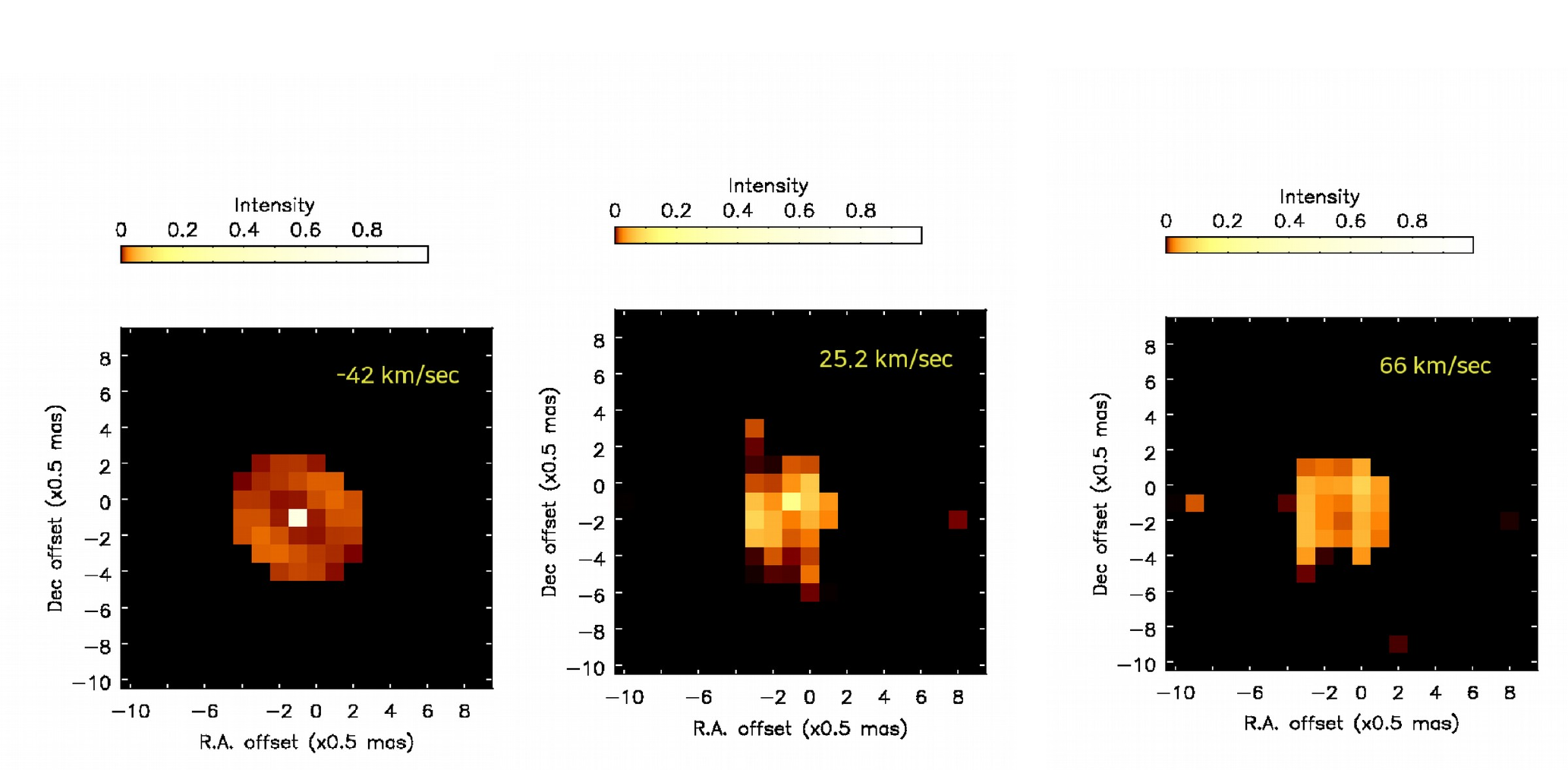} \\ \includegraphics[scale=1,trim={30 600 30 50},clip=true]{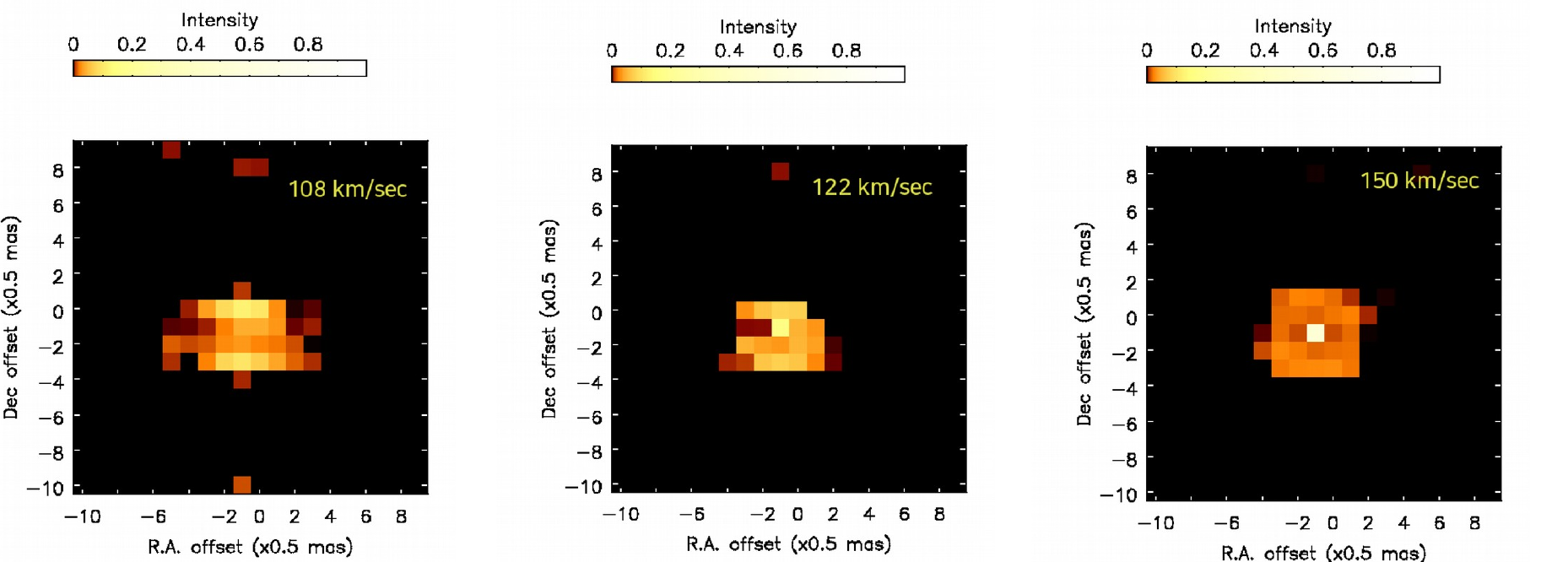} \\
\end{center}
\caption{Reconstructed channel velocity maps of the Br$\gamma$ emission towards IRAS 17163. The corresponding velocities are [-42, 25, 66, 108, 122, 150] kms$^{-1}$ (LSR).}
\label{fig:channel_imageBrg}
\end{figure*}

For the channel reconstruction maps of Br$\gamma$ we followed the same process as in Sect.~\ref{recon} but for the velocity channels at -42 (absorption component), 25, 66, 108, 122 and 150 kms$^{-1}$ (Fig.~\ref{fig:channel_imageBrg}). The channel maps reveal differences in morphology. In particular, at velocities $<$ 100 kms$^{-1}$ the emission shows 2 distinct emitting clumps in W-E, while for velocities $>$ 100 kms$^{-1}$ there is a change of 90 degrees in orientation. This could indicate an outflowing material in different velocities and orientations, which may all together lead to a more spherical and symmetric emission. Higher spectral and spatial resolution observations are required to confirm the observed differences among the different velocities, and will lead to a proper investigation of the substructures. 


\subsection{Combined dataset}
\label{bsmem}

\begin{figure*}[ht]
\begin{center}  
\includegraphics[scale=0.5]{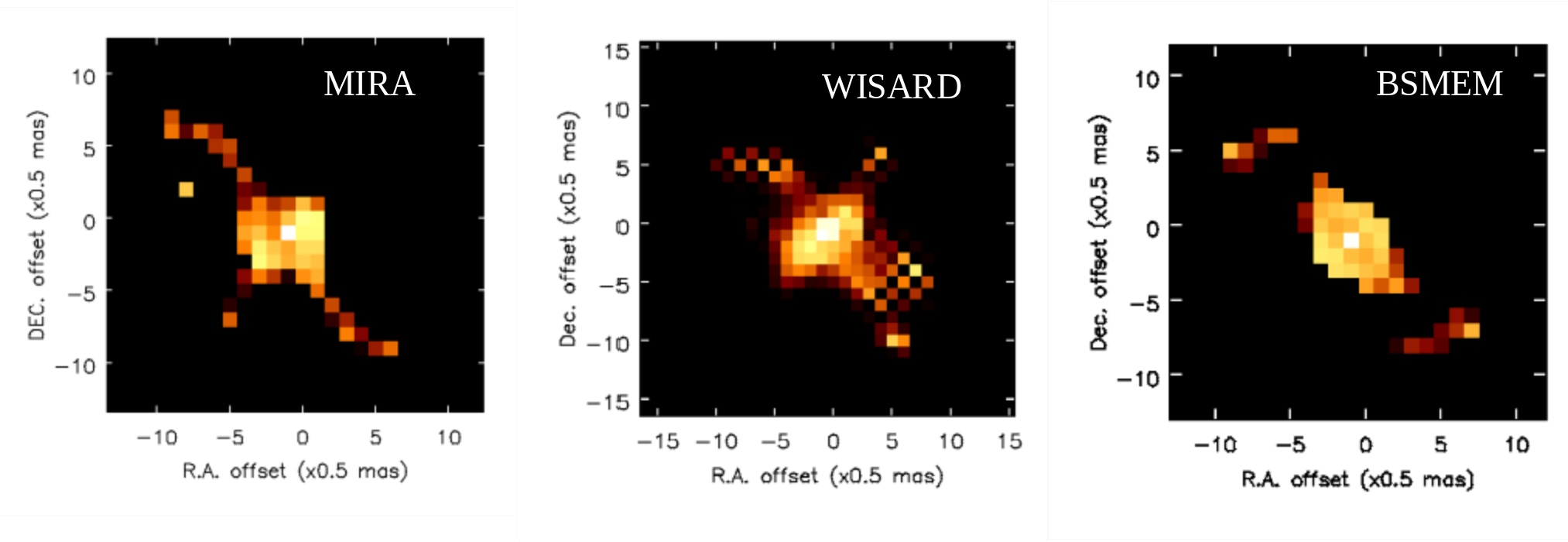} 
\end{center}
\caption{Image reconstruction of the Br$\gamma$ emission for the combined dataset (AMBER+GRAVITY) and for three different software packages as appear from left to right: MIRA, WISARD and BSMEM. The image is algorithm dependent, but they all show a similar global orientation of the emission. The resulted image using BSMEM shows an elongated structure around R.A. offset, Dec offset = (-5, 5), (-5, 5), which is perpendicular to that of the other two software packages. The observed difference is due to the different regularisers and optimisation process available for each algorithm.}
\label{fig:imageBrgcont_soft}
\end{figure*}

In addition to Sect.~\ref{recon} we combined the AMBER and the GRAVITY dataset and followed the same process for the Br$\gamma$ and continuum. Br$\gamma$ shows time variations and its morphology is expected to be affected. Therefore although the reconstructed image of the combined dataset is a result of a more complete uv-coverage compared to the GRAVITY dataset alone, the image reconstruction is less robust due to the observed flux and size variations of the Br$\gamma$ emission at the different epochs (see also Fig.~\ref{fig:Na IBrg2}). Given the above, the reconstructed image of the combined dataset (Fig.~\ref{fig:imageBrgcont_soft}) naturally differs from that of the GRAVITY dataset (Fig.~\ref{fig:imageBrgcont}). 

A thorough comparison with other software packages was crucial in order to be confident that what the image shows are real structures and not artifacts. For this purpose in addition to MIRA we used BSMEM \citep{Baron2008} and WISARD \citep{Meimon2004,Meimon2008}. We present the resulted images of the Br$\gamma$ spatial distribution for the combined dataset in Fig.~\ref{fig:imageBrgcont_soft}. The reconstructed images show similarities regarding the asymmetric and extended nature of the Br$\gamma$. In particular all algorithms reveal features in the NW--SE orientation. The resulted image using BSMEM shows an elongated central emission perpendicular to that of the other two software packages. The observed differences that can be seen in the detailed, pixel sized, structures in the combined dataset, are due to the different ways the algorithms treat the interferometric observables, such as the different available regularisers and optimisation process (i.e. gradient vs global) per algorithm. Therefore, the observed differences among the algorithms are software-related artifacts. We conclude that the observed elongation in NW-SE orientation is real and indicative of an asymmetric recent mass-loss, but how ``thin'' or ``thick'' this emission is, is something we cannot constrain with the current dataset. 

\end{appendix}
\end{document}